\documentclass[twocolumn,pra,aps,superscriptaddress,showpacs]{revtex4-1}

\usepackage{dcolumn}

\usepackage{mathptmx}   
\usepackage{bm}         

\usepackage[english]{babel}

\usepackage[pdftex]{graphicx} 
\usepackage{float}
\usepackage{epstopdf}
\usepackage{color}
\definecolor{mygrey}{gray}{0.45}
\definecolor{myblue}{rgb}{0.2,0.2,0.8}
\definecolor{myzard}{cmyk}{0,0,0.05,0}
\definecolor{mywhite}{rgb}{1,1,1}
\definecolor{myred}{rgb}{1,0.,0.3}
\usepackage[colorlinks=true,citecolor=myblue,linkcolor=myred,hidelinks=true]{hyperref}

\usepackage{latexsym}
\usepackage{amsmath,amssymb,amsfonts,amsbsy,mathtools}
\usepackage{upgreek}    
\usepackage{mathrsfs}   
\usepackage{cancel}     
\usepackage{mathdots}   
\usepackage{setspace}   
\usepackage{dsfont}     

\usepackage{enumerate}  

\usepackage{tikz}
\usetikzlibrary{matrix}
\usepackage{float}

\usepackage{appendix}

\usepackage{natbib}

\newcommand{\bra}[1]{\left\langle #1\right|}
\newcommand{\ket}[1]{\left| #1\right\rangle}
\newcommand{\braket}[2]{\langle #1|#2\rangle}

\newcommand\kk{\mathbf{k}}
\newcommand\rr{\mathbf{r}}
\newcommand\nn{\mathbf{n}}

\newcommand\bc[1]{a_{#1}^\dagger}
\newcommand\bd[1]{a_{#1}}
\newcommand\ux{\hat{\mathbf{u}}_x}
\newcommand\uy{\hat{\mathbf{u}}_y}
\newcommand\uz{\hat{\mathbf{u}}_z}

\newcommand\Nem{N_{\mathrm{em}}}    
\newcommand\weylfreq{\omega_W}      
\newcommand\stghop{h}               
\newcommand\stgmas{m}               
\newcommand\g{g}                    
\newcommand\BZ{\mathrm{BZ}}         

\begin{document}

\title{Light-matter interactions near photonic Weyl points}
\author{I\~naki Garc\'{i}a-Elcano}
\affiliation{Departamento de F\'{i}sica Te\'{o}rica de la Materia Condensada  and Condensed Matter Physics Center (IFIMAC), Universidad Aut\'{o}noma de Madrid, E-28049 Madrid, Spain}
\author{Jorge Bravo-Abad}
\email{jorge.bravo@uam.es}
\affiliation{Departamento de F\'{i}sica Te\'{o}rica de la Materia Condensada  and Condensed Matter Physics Center (IFIMAC), Universidad Aut\'{o}noma de Madrid, E-28049 Madrid, Spain}
\author{Alejandro Gonz\'alez-Tudela}
\email{a.gonzalez.tudela@csic.es}
\affiliation{Instituto de F\'{i}sica Fundamental IFF-CSIC, Calle Serrano 113b, Madrid 28006, Spain}

\begin{abstract}

Weyl photons appear when two three-dimensional photonic bands with linear dispersion are degenerate at a single momentum point, labeled as Weyl point. These points have remarkable properties such as being robust topological monopoles of Berry curvature as well as an associated vanishing density of states. In this work, we report on a systematic theoretical study of the
quantum optical consequences of such Weyl photons. First, we analyze the dynamics of a single quantum emitter coupled to a Weyl photonic bath as a function of its detuning with respect to the Weyl point and study the corrresponding emission patterns, using both perturbative and exact treatments. Our calculations show an asymmetric dynamical behavior when the emitter is detuned away from the Weyl frequency, as well as different regimes of highly collimated emission, which ultimately translate in a variety of directional collective decays. Besides, we find that the incorporation of staggered mass and hopping terms in the bath Hamiltonian both enriches the observed phenomenology and increases the tunability of the interaction. Finally, we analyze the competition between the coherent and dissipative components of the dynamics for the case of two emitters and derive the conditions under which an effective interacting spin model description is valid.

\end{abstract}

\maketitle

\section{Introduction}

The connection between the quantized Hall conductance displayed by  a two-dimensional electron gas subjected to a strong magnetic field~\cite{klitzing1980a} and its topological foundations~\cite{thouless1982a,kohmoto1985a} gave rise to the notion of topological phases of matter~\cite{haldane1988a,kane2005a}. This concept has now permeated a broad range of disciplines beyond solid state physics, including photonics~\cite{lu2014a,ozawa2019a}, mechanics~\cite{Huber2016a}, acoustics~\cite{x.zhang2018a}, and ultra cold atomic gases~\cite{cooper2019a}. In particular, the extension of these ideas to the photonics realm was first conceived as a natural generalization of the chiral edge states obtained within the quantum Hall effect~\cite{haldane2008a}. These chiral edge states were experimentally demonstrated in a magneto-optical photonic crystal a year after the being suggested~\cite{wang2009a}, and since then, a great amount of photonic realizations have been proposed to mimic the topological phenomena displayed by condensed matter models, including coupled waveguides~\cite{rechtsman2013a}, arrays of optical resonators~\cite{hafezi2011a,hafezi2013a} or  bianisotropic~\cite{khanikaev2013a} and chiral~\cite{w.gao2015a} metamaterials, among a number of other important advances using a variety of platforms and frequency regimes
\cite{fang2012a,umucalilar2012a,kraus2012a,ozawa2014a,price2014a,jacqmin2014a,kapit2014a,bliokh2015a,price2015a,aidelsburger2015a,l.h.wu2015a,ningyuan2015a,jin2016a,mechelen2016a,q.lin2016a,lu2016a,w.gao2016a,wj.chen2016a,m.xiao2016a,goldman2016a,schine2016a,iadecola2016a,anderson2016a,jin2017a,jw.dong2017a,maczewsky2017a,mukherjee2017a,b.yang2017a,slobozhanyuk2017a,j.noh2017a,milicevic2017a,ozawa2017a,st-jean2017a,wimmer2017a,x.piao2018a,j.noh2018a,q.lin2018a,b.yang2018a,mukherjee2018a,zilberberg2018a,h.jia2019a,mittal2019a,y.yang2019a,d.wang2019a}. Of special interest for this work are topological photonic structures that support the so-called Weyl points~\cite{lu2015a}. These are points in reciprocal space in which two linearly dispersive bands touch creating a singular band-gap, much in the same way as occurs in Dirac two-dimensional systems~\cite{castroneto09a}. However, a key difference with respect to them is that Weyl points are topologically protected: a Weyl point can only be annihilated when it meets another Weyl point with opposite chirality~\cite{armitage18a}, providing them important advantages such as robustness to disorder \cite{buchhold18a,buchhold18b}.

Most of the above referred topological photonic implementations have so far operated in the linear regime, that is, when the photon-photon interactions are very weak and do not play a significant role in the emerging phenomena. The interplay between topology and interactions is known to lead to strongly-correlated phenomena in fermionic models, such as fractional quantum Hall states~\cite{stormer99a}, that will be desirable to export as well into the photonics realm. A way to incorporate such interactions in these topological photonic systems is by coupling them to non-linear elements, like quantum emitters. The coupling of emitters to these topological photonic structures is challenging but has already been realized in pioneering experiments with photonic crystals~\cite{barik2016a,barik2018a} and superconducting circuits~\cite{E.kim2020a}. Besides, novel approaches involving quantum metasurfaces~\cite{perczel2017a,bettles2017a,perczel2020a,perczel2020b,masson2019atomic,patti2020controlling,rui2020subradiant} and innovative techniques based on matter-waves in state dependent optical lattices~\cite{krinner2018a} foresee even more implementations of these models in the near-future. All these experimental advances are consequently triggering an exciting --though still mostly unexplored-- area of research that investigates the individual and collective quantum optical phenomena which occur when one or more emitters interact with 1D~\cite{bello2019a,leonforte2020a}, 2D~\cite{leonforte2020a,bernardis2020a}, and 3D~\cite{garcia-elcano2020a,ying2019a} topological photonic reservoirs. In this context, we have recently showed how the emergence of a light-matter bound state in 3D photonic Weyl systems enables coherent, tunable and robust power-law interactions among quantum emitters~\cite{garcia-elcano2020a}, paving the way towards more robust long-distance entanglement protocols or quantum simulation implementations for studying long-range interacting systems. However, a systematic theoretical analysis of all the quantum optical phenomena emerging from these platforms is still missing.

In this work, we present an in-depth analysis of the quantum optical behavior displayed by quantum emitters interacting with the bulk modes of a Weyl photonic environment. In particular, \emph{(i)} We show how the interactions between emitters can be tuned by incorporating both a staggered mass and a staggered hopping term in the bath Hamiltonian; \emph{(ii)} We study the exact quantum dynamics of the single and two-emitter cases, finding an asymmetric dynamical behavior when the emitters are detuned away from the Weyl frequency, and also atypical incomplete coherent exchange oscillations when the two emitters couple to different sublattices; \emph{(iii)} We also study the emission patterns when the emitter's energy lies within the band, uncovering the emergence of highly-collimated emission for certain frequency regions; \emph{(iv)} Finally, we discuss the competition between the dissipative and coherent components of the dynamics in the many-emitter configuration, allowing us to find the parameter regimes where an effective spin description based on perturbative (Markovian) treatments is valid.

The rest of the manuscript is organized as follows. In Section~\ref{System} we introduce the system under study, whereas in Section~\ref{Theoretical framework} we account for the analytical and numerical strategies employed throughout the manuscript. In Section~\ref{Single emitter}, we investigate the time evolution of a single emitter coupled to the bath using both a perturbative (Markovian) and an exact treatment. In addition, to the single-emitter dynamics, we study the shape of the radiative emission patterns and of the emergent bound states of the system. Section~\ref{Two emitters} is devoted to the analysis of the dynamical behavior of the two emitters case and is divided in two subsections. First, in Section~\ref{Same sublattice} we consider a configuration in which the two emitters are coupled to sites belonging to the same sublattice, there we include a discussion concerning the competition between the coherent and the dissipative components of the system's dynamics. Second, in Section~\ref{Different sublattice}, we consider the case in which the pair of emitters are coupled to sites belonging to different sublattices, which leads to qualitatively different behaviour. Then, in Section~\ref{Many emitters} we report on the effective spin model description that can be obtained by adiabatically eliminating the photonic degrees of freedom in the perturbative limit. Finally, in Section~\ref{Conclusions} we sum up the most interesting results of our work and point out potential directions of further research.

\section{System} \label{System}

The Hamiltonian of the studied system contains three contributions: $H=H_{M}+H_{B}+H_{I}$, where $H_{M}$ and $H_{B}$ account for the matter and photonic degrees of freedom respectively, whereas $H_{I}$ describes the interplay between them.
The most general implementation of our set up includes $\Nem$ emitters modeled as a collection of two-levels systems with transition frequency $\omega_{j}$. The associated dynamics (using $hbar=1$) is given by:
\begin{equation}
    H_{M}=\sum_{j=1}^{\Nem}\omega_{j}\sigma_{ee}^{j}
\end{equation}
where $\sigma_{\mu\nu}^{j}=\ket{\mu_{j}}\bra{\nu_{j}}$ $\left(\mu,\nu=g,e\right)$ are the atomic operators of the $j$-th quantum emitter which can be either on its ground, $\ket{g_{j}}$, or its excited, $\ket{e_{j}}$, state. 

For the structured bath, $H_{B}$, we start with a regular cubic arrangement of localized bosonic modes spaced by the lattice parameter $a$ (which we use as the unit of length henceforth). 
The Weyl points emerge as a consequence of breaking inversion symmetry after including an alternating change of sign for the hopping matrix elements along the x and z directions \cite{dubcek2015a,roy2018a}. 
We also incorporate an alternating on-site energy off-set which we refer to as the staggered mass term, $\stgmas$, and a staggered hopping term along the y direction, $\stghop$. These are used as tunable parameters that will allow us to explore the robustness and tunability of the Weyl points. The particular way how to create such tunable parameters will depend on the implementation chosen, which lies beyond the scope of this work. The bath Hamiltonian can then be described by a tight-binding model with nearest-neighbor interactions and hopping rate $J$. In real space $H_B$ reads:
\begin{equation}\label{eq:Real space Hamiltoninan}
H_B =
\sum_\rr \mathcal{E}_\rr \bc{\rr} \bd{\rr}
-\sum_{i=x,y,z} 
\left[\mathcal{J}_i \, \bc{\rr} \bd{\rr+\hat{\mathbf{u}}_i} + \mathrm{H.c.}\right]
\end{equation}
where $\bc{\rr}/\bd{\rr}$ stands for the bosonic creator/annihilation operator at site $\rr=(x,y,z)$, $\hat{\mathbf{u}}_i$ with $i=x,y,x$ denote the unitary vectors along the three principal axes, and we have defined:
\begin{itemize}
    \item On-site energy: $\mathcal{E}_\rr = \weylfreq-\stgmas(-1)^{x+y}$
    \item Hopping along $\ux$, $\uz$: $\mathcal{J}_{x,z} = J(-1)^{x+y}$
    \item Hopping along $\uy$: $\mathcal{J}_y = 
    \frac{1}{2}[(J+\stghop)+(-1)^{x+y}(J-\stghop)]$
\end{itemize}
Note that the on-site energy of the bosonic modes provided that $\stgmas=0$ has been chosen to be $\weylfreq$, which we identify as the Weyl frequency.

\begin{figure}[tb]
	\centering
	\includegraphics[width=\columnwidth]{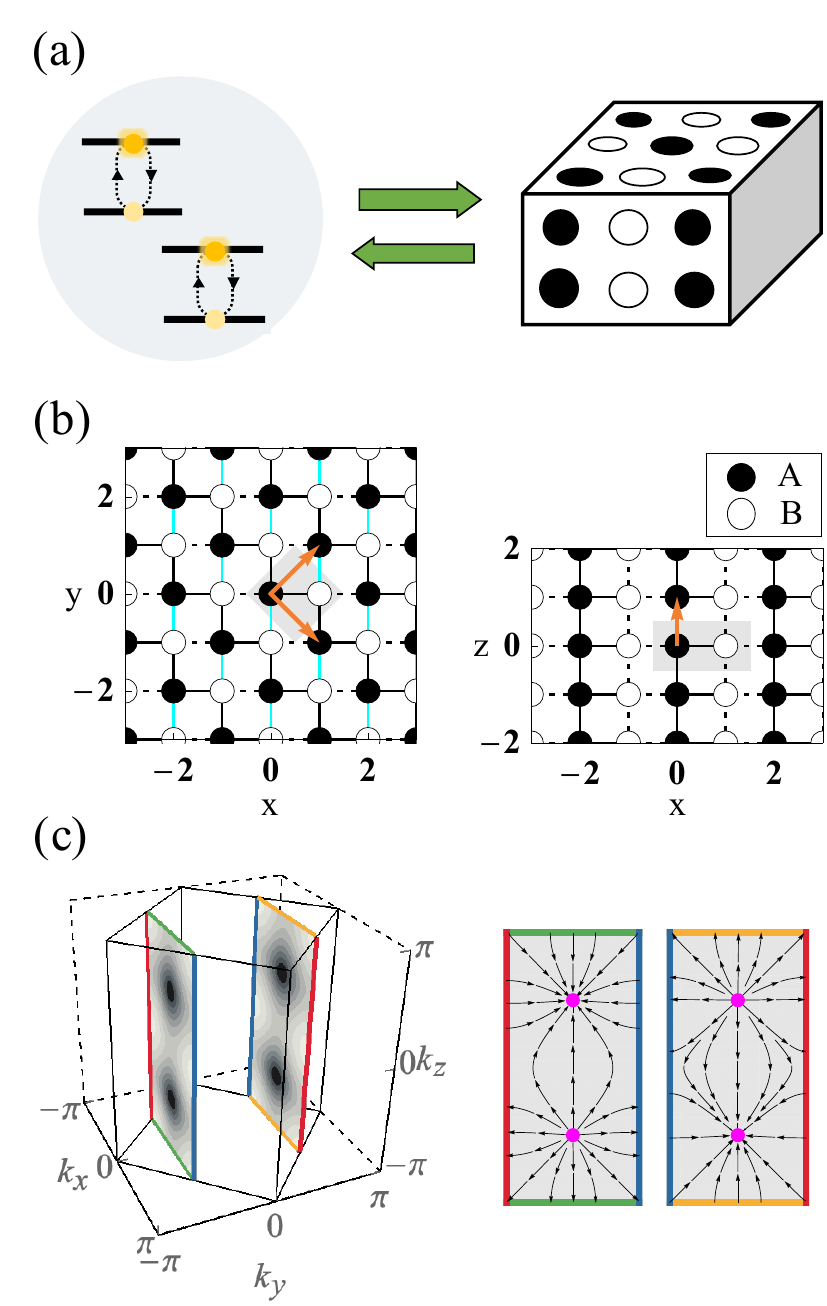}
	\caption{
	   (a) Schematics of the analyzed system.
	   (b) Top and side view of the three-dimensional lattice model mimicking the Weyl environment. Solid/dashed black lines correspond to a hopping term of magnitude $J$ and phase $0$/$\pi$ whereas cyan solid lines stands for the staggered hopping term. Orange arrows indicate the primitive vectors and the shadow gray area encloses the unit cell. 
	   (c) First Brillouin zone of the investigated lattice model. The associated dispersion relation is projected over the $k_y=\pm\pi/2$ planes. The inset show a detailed view these planes in which the position of the Weyl points (magenta dots) and the Berry curvature (vector map) is depicted for the case $m=0$ and $h/J=1$.
	   \vspace{-0.3cm}}
	\label{fig:Fig1}
\end{figure}

The bath Hamiltonian is invariant under discrete translation of $\mathbf{R}_{\nn}=\sum_{i} n_{i}\mathbf{a}_{i}$, where $n_{i}\in\mathbb{Z}\;\forall i$ and the corresponding primitive vectors, $\mathbf{a}_{i}$, are displayed as orange arrows in Fig.\ref{fig:Fig1}(b). The lattice unit cell is formed by a pair of nonequivalent sites belonging to two different sublattices that we label as $A$ and $B$. Imposing periodic boundary conditions allows us to introduce reciprocal space where the bipartite nature of the lattice shows up as a pseudo-spin degree of freedom. To do that we define the Bloch operator:
\begin{equation}
    \bc{\kk,A/B}=\frac{1}{\sqrt{N_{A/B}}}\sum_{\rr\in A/B} e^{i\kk\rr} \bc{\rr}
\end{equation}
where the summand runs over the $N_{A/B}$ sites belonging to the $A/B$ sublattice and $\kk$ denotes a vector of the first Brillouin zone (see Fig.\ref{fig:Fig1}(c)). By inverting the Bloch operator and introducing it on Eq.~\eqref{eq:Real space Hamiltoninan}, we can rewrite the bath Hamiltonian as follows:
\begin{equation}
    H_B = \sum_{\kk} A_{\kk}^\dagger \, \bar{H}(\kk) \, A_{\kk}
\end{equation}
where $A_{\kk}=(\bd{\kk,A} \; \bd{\kk,B})^T$ and $\bar{H}(\kk)$ can be expressed in terms of the Pauli matrices, $\bm{\sigma}=(\sigma_x,\sigma_y,\sigma_z)$, such that:
\begin{equation}
    \bar{H}(\kk)=\omega_W\mathds{1}-\mathbf{d(\kk)}\bm{\sigma}
\end{equation}
with:
\begin{align}
    d_x = & (J+\stghop)\cos(k_y)
    \\
    d_y = & -2J\sin(k_x) - (J-\stghop)\sin(k_y)
    \\
    d_z = & \stgmas + 2J\cos(k_z)
\end{align}
Finally, we consider a unitary transformation that diagonalizes $\bar{H}(\kk)$, leading to a compact form of the bath Hamiltonian which reads:
\begin{equation}
    H_B=\sum_{\kk,\gamma}E_{\gamma}(\kk)\bc{\kk,\gamma}\bd{\kk,\gamma}
\end{equation}
where $\gamma$ runs over the upper and lower bands and  $E_{U/L}(\kk)=\omega_W\pm\omega(\kk)$ with $\omega(\kk)=|d(\kk)|$ (see Fig.~\ref{fig:Fig1}c for an two-dimensional projection of these energy bands in the Brillouin zone). It must be noticed that the dispersion relation is composed by two bands which are symmetric with respect to $\omega_{W}$, irrespective of the bath parameters. 
The variation of both the staggered mass and the staggered hopping terms gives rise to a smooth modification of the Weyl points' positions in reciprocal space. For $\left|\stgmas/J\right|>2$ a gap is opened, and the same occurs whether $\stghop/J<-1$ or $\stghop/J>3$. It is worth noting that for $\stghop/J=-1$ we have $\mathcal{J}_y=J(-1)^{x+y}$, so that a staggered $\pi$ phase for the hopping along the $y$ direction is assumed. For $\stghop/J=3$, the hopping phase along the $y$ direction remains unchanged but the corresponding amplitude displays an alternating value. In order to recover a uniform hopping, both in phase and amplitude, along the $y$ direction we must set $h/J=1$.

The interaction between the quantum emitters and the bosonic modes of the bath is assumed to be local and it features a constant coupling strength, $\g$. Additionally, we neglect counter-rotating processes which is a faithful assumption within the optical regime where $\omega_{j}\approx\omega_{W}\gg \g$. In real space the corresponding Hamiltonian reads:
\begin{equation}
    H_{I}= \g \sum_{j=1}^{\Nem}\left(\bc{\rr_j}\sigma_{ge}^{j}+\mathrm{H.c.}\right)
\end{equation}
where $\rr_j$ denotes the position of the $j$-th quantum emitter. This interacting term has the form of the well known Jaynes Cummings Hamiltonian which is traditionally employed to model light-matter interactions in photonic crystals waveguides~\cite{goban13a,thompson13a}, in state dependent optical lattices~\cite{devega08a,krinner2018a}, as well as microwave resonators coupled to superconducting circuits~\cite{E.kim2020a}, as long as they do not enter in the ultra-strong coupling limit~\cite{kockum2019ultrastrong}.

Once the total Hamiltonian has been presented it is worth noting that it commutes with the number of excitations' operator:
$N_{exc} = \sum_{j=1}^{\Nem}\sigma_{ee}^{j}+\sum_{\kk,\gamma}\bc{\kk,\gamma}\bd{\kk,\gamma}$. Therefore, each excitation subspace can be treated separately. Also, it is important to realise that the dynamics of the quantum emitters can be much more easily analysed within a rotating frame in which the oscillations associated to the value of the Weyl frequency are factorized:
\begin{equation}
    e^{-iHt}=e^{-i\omega_{W}N_{exc}t}e^{-i\left(H_{0}+H_{I}\right)t}\,,
\end{equation}
where we have introduced $H_{0}$ as:
\begin{equation}
	H_{0} = \sum_{j}^{\Nem}\Delta_{j}\sigma_{ee}^{j}+\sum_{\kk}\omega(\kk)
                \left(\bc{\kk,U}\bd{\kk,U}-\bc{\kk,L}\bd{\kk,L}\right)\,,
\end{equation}
and where we have defined the detuning of the $j$-th emitter with respect to the Weyl frequency $\Delta_{j}\equiv\omega_{j}-\omega_{W}$.

\section{Theoretical framework} \label{Theoretical framework}

The dynamics of a collection of quantum emitters interacting with a Weyl reservoir can be equivalently obtained using either analytical or numerical strategies.

To compute the emitters' dynamics analytically we use the so-called \emph{resolvent operator method} \cite{cohen_book92a}. This technique benefits from identifying a small subset of states which play a relevant role in the investigated physical problem. In our case, most of the phenomena that we will analyze can be restricted to the single-excitation subspace, which allows us to write down the global system wavefunction as:
\begin{equation}\label{eq:overall-system-state}
    \ket{\Psi(t)}=\left[\sum_{j=1}^{\Nem}C_{j}(t)\sigma_{eg}^{j}+\sum_{\kk,\gamma}C_{\kk,\gamma}(t)\bc{\kk,\gamma}\right]\ket{\Psi_0}
\end{equation}
where $j$ runs over emitters, $\kk$ runs over vectors belonging to the first Brillouin zone and $\gamma$ runs over different bands. Here, we have implicitly divided the Hilbert space of the problem in two sets: on the one hand we have the states spanning the photonic degrees of freedom, $\ket{\phi_{\kk,\gamma}}=\bc{\kk,\gamma}\ket{\Psi_0}$, and on the other hand the states spanning the emitter degrees of freedom, $\ket{\phi_{j}}=\sigma_{eg}^{j}\ket{\Psi_0}$, where $\ket{\Psi_0}\equiv\ket{g_{1}\cdots g_{\Nem};\mathrm{vac}}$ denotes the overall vacuum state.

The dynamics of the $j$-th quantum emitter is characterized by the time evolution of its excited state's population, $\left|C_{j}(t)\right|^{2}$, where $C_{j}(t)$ is the projection of the overall-system wave function at arbitrary time $t$ over any specific emitter state: 
\begin{equation}
    C_{j}(t) = \braket{\phi_j}{\Psi(t)}=\bra{\phi_j}U(t)\ket{\Psi(0)}\,,
\end{equation}
being $U(t)=e^{-i\left(H_{0}+H_{I}\right)t}$ the unitary time evolution operator, and where we need to specify an initial condition. In this work, we always consider situations in which the initial state of the system contains no excitations within the bath. Accordingly, we focus on calculating the matrix elements of the evolution operator (i.e. the transition probabilities) given by $ U_{jj^{\prime}}(t)=\bra{\phi_j}U(t)\ket{\phi_{j^{\prime}}}$. These are connected to the matrix elements of the propagator (or resolvent) of the system, $G_{jj^{\prime}}(z)=\bra{\phi_{j}}G(z)\ket{\phi_{j^{\prime}}}$, by complex integration as follows:
\begin{equation}\label{eq:Propagator inversion}
    U_{jj^{\prime}}(t)=\int_{C_{+}+C_{-}}\frac{dz}{2\pi i}G_{jj^{\prime}}(z)\,e^{-izt},
\end{equation}
where $C_{+}/C_{-}$ accounts for the retarded/advanced contribution to the propagator, which vanishes for negative/positive times. Note that further information concerning the calculation of Eq.~\eqref{eq:Propagator inversion} can be found in appendix~\ref{appendix:inversion}.

For a fixed number of emitters, the matrix elements of the propagator, $G_{jj^{\prime}}(z)$, can be computed solving the system of equations defined by:
\begin{equation} \label{eq:Propagator system of eqs}
\sum_{n=1}^{\Nem}\left[z\delta_{jn}-\epsilon_{jn}-\Sigma_{jn}\right]G_{nj^{\prime}}=\delta_{jj^{\prime}}
\end{equation}
where $\epsilon_{jn}=\bra{\phi_j}H_0\ket{\phi_n}$ and where we have identified the matrix elements of the problem's self-energy as:
\begin{equation}\label{eq:Self-energy matrix elements}
    \Sigma_{jj^{\prime}}(z)=
    \sum_{\kk,\gamma}
    \frac{
    \bra{\phi_{j}}H_{I}\ket{\phi_{\kk,\gamma}}
    \bra{\phi_{\kk,\gamma}}H_{I}\ket{\phi_{j^{\prime}}}}
    {z-\bra{\phi_{\kk,\gamma}}H_0\ket{\phi_{\kk,\gamma}}}
\end{equation}
Since the considered Hamiltonian only includes excitation conserving terms, the particular form of the self energy matrix elements given by Eq.~\eqref{eq:Self-energy matrix elements} is exact. More complicated models, e.g. taking into account multiple photons, will involve further summations with higher order denominators such that, in order to perform any concrete calculation, one usually needs to truncate the series.
However, even for the simple case of a Hamiltonian operating in the single particle subspace, calculating the matrix elements of the self-energy operator might be a challenging task. It must be noticed that $\Sigma_{jj^{\prime}}(z)$ depends on the configuration of the studied system which is determined by the parameters characterizing the bath (i.e. staggered mass and hopping terms), the light-matter coupling constant, the number and position of the considered emitters, and, in our case, also the sublattice to which they are coupled to.

An alternative strategy to obtain the emitters' dynamics consists in solving the Schr\"{o}dinger equation for the total Hamiltonian numerically. This can be done computing the product of the exponential of the Hamiltonian's matrix with a vector representing the initial state of the system in a convenient basis of localized bosonic sites. For that we make use of the algorithm developed in \cite{al-mohy2011a}. A more detailed description of the employed numerical strategy can be found on appendix~\ref{appendix:numerics}.

\section{Single emitter: dynamics, emission patterns, and bound-states} \label{Single emitter}

\subsection{Dynamics}

As a starting point let us consider a single quantum emitter coupled to the Weyl environment. We assume that the system is prepared with the emitter in its uppermost state and no exciations in the bath, i.e.  $\ket{\Psi(0)}=\ket{e;\mathrm{vac}}$. Within the resolvent approach, the excited state population of the emitter after time $t$, $\left|U_{11}(t)\right|^2\equiv\left|C_e(t)\right|^2$, can be obtained inverting the corresponding propagator's matrix element:
\begin{equation}\label{eq:G_e(z)}
    G_e(z)=\frac{1}{z-\Delta-\Sigma_e(z)}
\end{equation}
The self-energy of the single emitter case in the thermodynamic limit reads:
\begin{equation}\label{eq:Single emitter Self-Energy (integral)}
\Sigma_{e}^{(A/B)}(z)=\g^{2}\iiint_{\BZ}\frac{d^{3}\kk}{V_{\BZ}}\;
\frac{z \mp [\stgmas+2J\cos(k_z)]}{z^{2}-\omega(\kk)^2}
\end{equation}
where the sign on the right hand side stands for the sublattice, marked by the superscript $(A/B)$, to which the emitter is coupled to. As anticipated, the self-energy matrix element depends on the particular configuration of the system, e.g., the term in brackets in the numerator arises as a consequence of the combined action of the inter-layer hopping (i.e. hopping along the $z$ direction) and a non-trivial value of the staggered mass term. Due to the symmetry of the integral, the latter term vanishes for the case $m=0$.

A standard way of attacking these problems is to consider perturbative treatments, such as the Markov approximation, which can be recovered within the resolvent operator formalism by neglecting the complex dependence of the self energy's matrix element and evaluating it at the emitter's detuning value $\Delta$, i.e.,  $\Sigma_{e}(\omega+i0^{+})\approx\Sigma_{e}(\Delta+i0^{+})$. This is generally a good approximation as long as the light-matter coupling, $\g$, is weak enough and that $\Delta$ does not match a van Hove singularity in the spectrum of the bath~\cite{Gonzalez-Tudela2017a}. Using that approximation, the propagator of the problem (see Eq.~\eqref{eq:G_e(z)}) has a particularly simple form which can be easily inverted, leading to the following temporal evolution for the $C_{e}(t)$ coefficient:
\begin{equation}
    C_{e}(t) \approx e^{-i\,\left[\Delta+\delta\omega_{e}(\Delta)-i\,\frac{\Gamma_{e}(\Delta)}{2}\right]\,t}
\end{equation}
Here $\delta\omega_{e}(\Delta) \equiv \mathrm{Re}\,\Sigma_{e}(\Delta+i0^{+})$ and $\Gamma_{e}(\Delta) \equiv -2\,\mathrm{Im}\,\Sigma_{e}(\Delta+i0^{+})$ can be respectively interpreted as an energy shift of the emitter's transition frequency and the exponential decay rate of the excited state population.
The later one is deeply connected with the density of states, that can be defined as the limit $z\rightarrow\Delta+i0^+$ of the function:
\begin{equation}
    D(z)= -\mathrm{Im}\frac{2}{\pi}
    \iiint_{BZ}\frac{d^3\kk}{V_{BZ}}\,\frac{z}{z^2-\omega(\kk)^2}
\end{equation}
which yields:
\begin{equation}
    D(\Delta)= 
    \iiint_{BZ}\frac{d^3\kk}{V_{BZ}}\,\left[\delta(\Delta-\omega(\kk)+\delta(\Delta+\omega(\kk)\right]
\end{equation}
For completeness, in this work we will complement and compare the Markovian predictions with the calculations of the dynamics based on the analytical or numerical methods discussed in section~\ref{Theoretical framework}.

Let us first study the most simple scenario where no staggered mass nor staggered hopping terms are considered i.e. $\stgmas=0$ and $\stghop/J=1$. These simplifications allow us to obtain a closed analytical expression for $\Sigma_{e}(z)$ whose form is independent of the sublattice to which the emitter is locally coupled to:
\begin{equation}
    \Sigma_{e}(z)=\frac{4\g^{2}}{\pi^{2}}\frac{z}{z^{2}-6J^{2}}
    \frac{1-9\xi^{4}}{(1-\xi)(1+3\xi)}\,K^{2}\left(m\right)
    \label{eq:Sigma_e(M=0)}
\end{equation}
where $K(z)$ stands for the complex elliptic integral of the first kind, and we have defined the following magnitude:
\begin{equation}
    m=\frac{16\xi^{3}}{(1-\xi)(1+3\xi)}
\end{equation}
with:
\begin{equation}
    \xi=\frac
    {\sqrt{1-\sqrt{1-\frac{1}{9}\left(\frac{6J^{2}}{z^{2}-6J^{2}}\right)^{2}}}}
    {\sqrt{1+\sqrt{1-\left(\frac{6J^{2}}{z^{2}-6J^{2}}\right)^{2}}}}
\end{equation}

\begin{figure}[tb]
	\centering
	\includegraphics[width=\columnwidth]{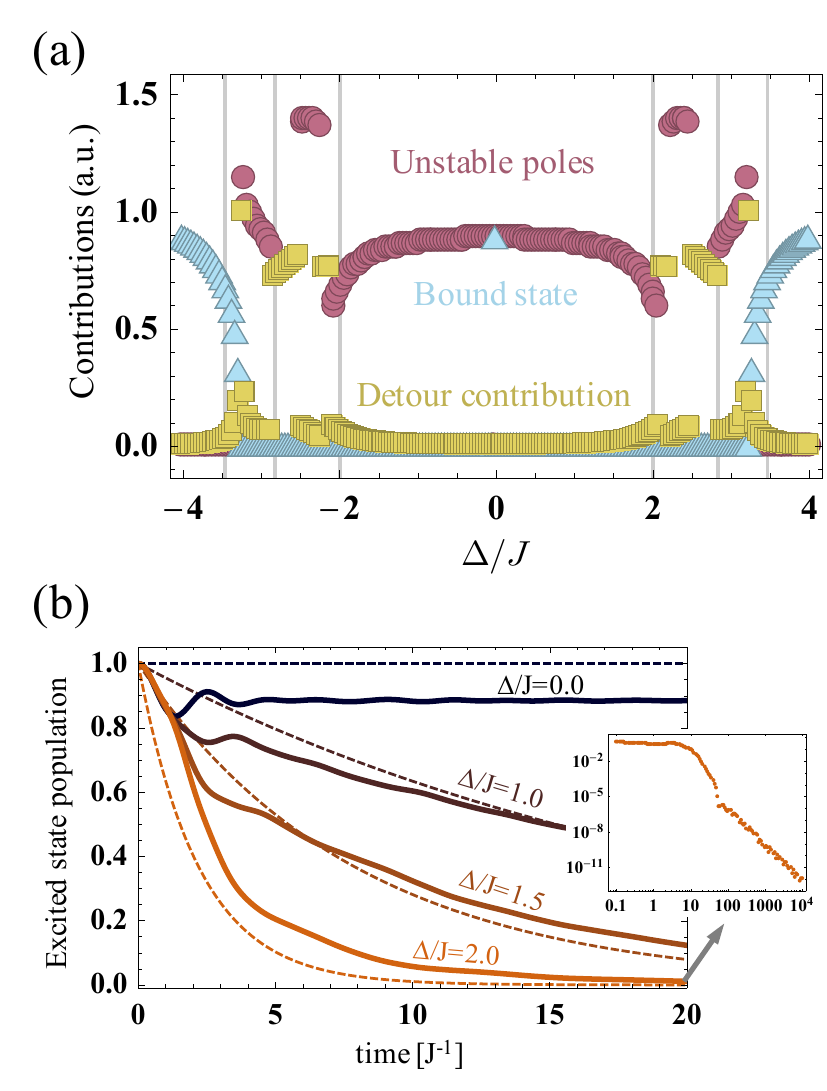}
	\caption{
	   Dynamics of the single emitter case for configuration in which $\stgmas=0$, $\stghop/J=1$ and the light-matter coupling is chosen to be $\g/J=0.5$.
	   (a) Contributions to $|C_e(0)|^2$ as a function of the detuning of the emitter with respect to the Weyl frequency. 
	   (b) Markovian (dashed lines) and non-Markovian (solid lines) prediction for the temporal evolution of the emitter's excited state population for various detuning values. The inset shows the long-time behavior of the emitter's excited population in a log-log scale for the case in which the emitter matches the van-Hove singularity corresponding to $\Delta/J=2$.
	   \vspace{-0.3cm}}
	\label{fig:Fig2}
\end{figure}

In appendix~\ref{appendix:Multi-valuation} we carry out a detailed examination of the multivalued branch structure of $\Sigma_e(z)$ which concludes with the identification of a single valued physical self-energy. After appropriate inversion of the corresponding propagator's matrix element (see details in appendix~\ref{appendix:inversion}), we recognize three main contributions to $C_{e}(t)$:
\begin{equation}
    C_{e}(t)=C_{e}^\mathrm{BS}(t)+C_{e}^\mathrm{UP}(t)+C_{e}^\mathrm{Detour}(t)
\end{equation}
whose relative relevance depends on the detuning of the emitter with respect to the Weyl frequency as numerically depicted in Fig.~\ref{fig:Fig2}(a) for a fixed coupling strength. These contributions are related to the mathematical structure of the associated propagator and we further discuss them in what follows:
\begin{itemize}
    \item \emph{Poles of $G_e(z)$.} The poles present in the lower half plane of the complex plane can be divided into unstable poles, $C_{e}^\mathrm{UP}(t)$, (magenta dots), featuring a negative imaginary part which determines the decay time of the excited state, and real poles, $C_{e}^\mathrm{BS}(t)$, (blue triangles), corresponding to infinite-lifetime bound states. The latter become the major contribution nearby spectral regions where the density of states is exactly zero. 
    
    \item \emph{Non-analytical points in $G_e(z)$.} The non-analytical structure of $\Sigma_e(z)$ at certain frequencies (denoted with vertical gray lines) forces detours in the contour of integration that also contribute to the dynamics, $C_{e}^\mathrm{Detour}(t)$, (yellow triangles). These are related to the appearance of van Hove singularities in the middle of the band-structure which, as it also occurs for other structured baths~\cite{Gonzalez-Tudela2017,gonzalez2018non}, lead to a strongly non-Markovian dynamical behavior for $t\neq 0$. This can be explicitly observed in the inset of Fig.~\ref{fig:Fig2}(b), where we plot the long time dynamics for the case in which the emitter is detuned to $\Delta/J=2$.
\end{itemize}

The temporal evolution of the emitter's excited state for the studied case is shown in Fig.~\ref{fig:Fig2}(b). Both, the approximate (dashed lines) and exact (solid lines) treatments are displayed together for several detuning values so that differences between the two approaches are made apparent. Particularly, we find that the exact results differ appreciably from the dynamics obtained within the Markovian approximation at short times or when $\Delta$ exactly matches a van Hove singularity (inset). Nonetheless, the most remarkable deviation from the Markovian prediction occurs when $\Delta=0$, i.e. when the transition frequency of the emitter coincide with the Weyl frequency. For that detuning the quantum emitter is subjected to the so-called fractional decay ---a stationary regime in which the two-level system undergoes an incomplete deexcitation. This behavior is physically ascribed to the emergence of a photon bound state, that we label as Weyl bound state~\cite{garcia-elcano2020a}, whose properties will be described more in detail in a later section.

\begin{figure}[t]
	\centering
	\includegraphics[width=\columnwidth]{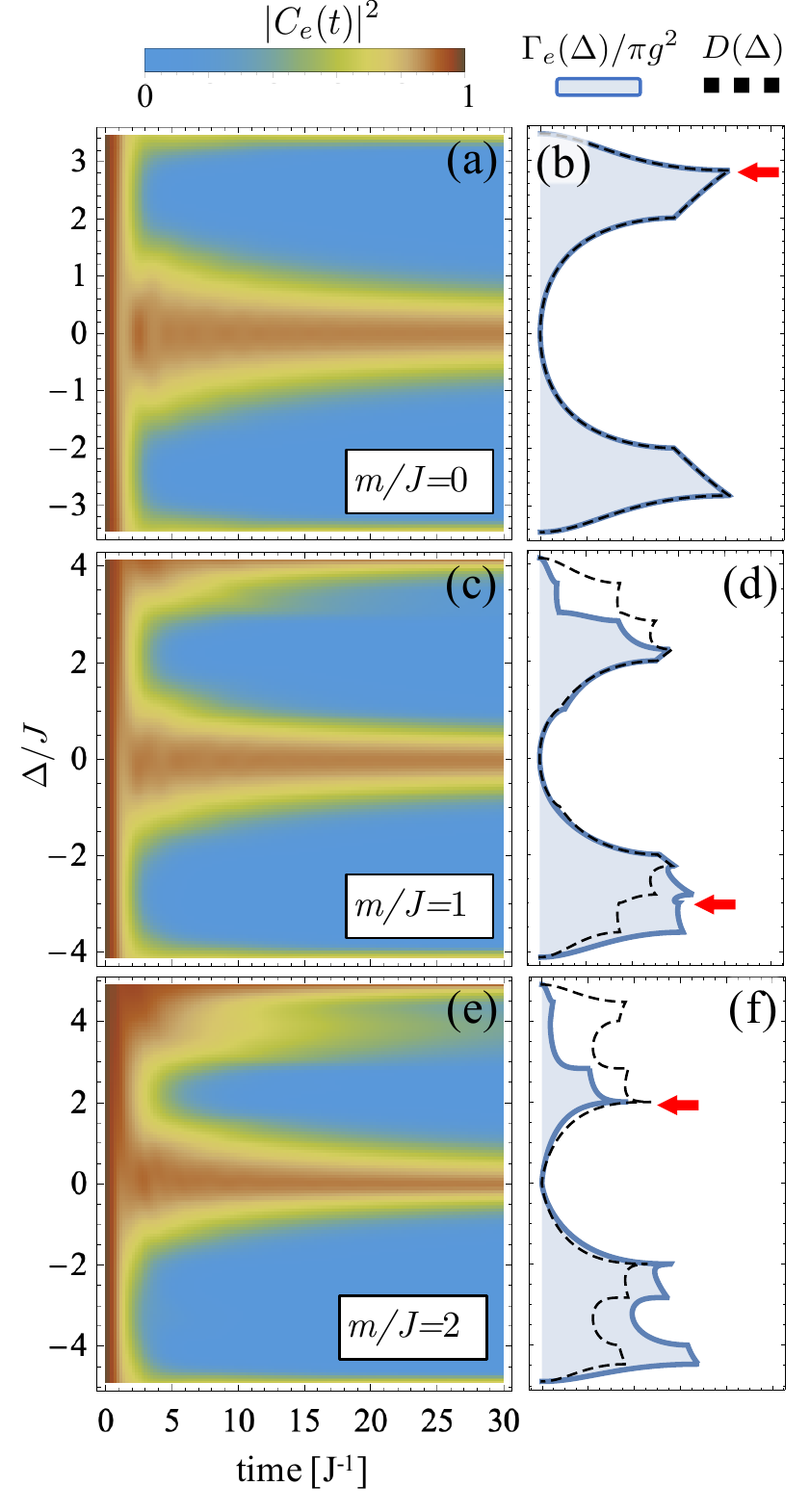}
	\caption{
	   (a,c,e) Excited state population of the emitter as a function of its detuning with respect to the Weyl frequency and time. 
	   (b,d,f) Comparison between the Markovian prediction for the decay rate and the density of states for several system configurations. Red arrows spot some selected van Hove singularities that are further explored in Fig.~\ref{fig:Fig4}. 
	   We restrict ourselves to configurations where no staggered hopping term is assumed ($\stghop/J=1$), but the staggered mass term is chosen to be $\stgmas/J=0$ (a,b), $\stgmas/J=1$ (c,d) and $\stgmas/J=2$(e,f). The light-matter coupling is $g/J=0.5$
	   \vspace{-0.5cm}}
	\label{fig:Fig3}
\end{figure}

Let us now consider the more general case of $\stgmas\neq 0$ but fixing the staggered hopping $\stghop/J=1$. The results are summarized in the color maps depicted in Fig.~\ref{fig:Fig3}(a,c,e), where we plot the temporal evolution of an emitter coupled to a site belonging to the $A$ sublattice as a function of $\Delta$ for several values of staggered mass term. It must be noticed that similar results are obtained when the emitter is locally coupled to a site belonging to the $B$ sublattice (not shown). We complement this figure with panels (b,d,f), where we plot both the associated density of states of the bath $D(\Delta)$ (in dashed black), together with the expected decay rate in the Markovian prediction $\Gamma_e(\Delta)$ (solid blue). From all these figures, we observe several features:
\begin{itemize}
    \item Near zero detuning the emitter features a vanishing decay rate which smoothly connects with the fractional decay regime associated to the emergence of the Weyl bound state.
    Consequently, we foresee that, within the appropriate time scale, the phenomenology associated to the emergence of such bound state will not present important changes under moderate perturbations of the emitter's detuning value.

    \item When the emitter's detuning lays inside the region where the decay rate $\Gamma_e(\Delta)$ acquires a significant value ($\Delta/J \gtrsim  1$), the emitter relaxes to the ground state in a much shorter time scale than in the previous case. For the cases in which $\stgmas\neq0$, it is interesting to note that, although the bath is characterized by a symmetric spectrum, the emitter displays distinct behavior when it is detuned towards the upper band or towards the lower band.  Namely, for increasing values of $\stgmas$ we observe that the emitter's population around the upper band-edge experiences a slower decay, and the other way around for negative frequencies. Such differences are already reflected in the asymmetric form of the Markovian prediction for the decay rate, $\Gamma_e(\Delta)$, and are coming from the non-trivial frequency-dependent structure of the coupling to the bath appearing when $\stgmas\neq 0$.
\end{itemize}

\subsection{Radiative emission patterns}

Beyond the temporal evolution of the emitter's population, it is also appealing to investigate the dynamics of the emitted excitations (i.e. dynamical behavior of the photonic degrees of freedom), since they will eventually be responsible of the collective decays appearing when more emitters are coupled to the bath~\cite{Gonzalez-Tudela2017a}. In general, these emission patterns will depend on the set of parameters characterizing the bath and the energy of the emitter. To illustrate that, we consider three particular configurations where the emitter's energy is tuned to one of the singularities present in $\Gamma_e(\Delta)$, since they are known to lead to exotic emission patterns in other structured bath configurations~\cite{Gonzalez-Tudela2017,gonzalez2018non}. Then, provided that the emitter is prepared in its excited state, we study the photonic component of the system at a certain de-excitation time. For that, we resort to our numerical method using a finite bath composed by a total of $2^{19}$ sites, and assuming that the emitter is coupled to a site belonging to the $A$ sublattice in the bulk of the material.

\begin{figure}[t]
	\centering
	\includegraphics[width=\columnwidth]{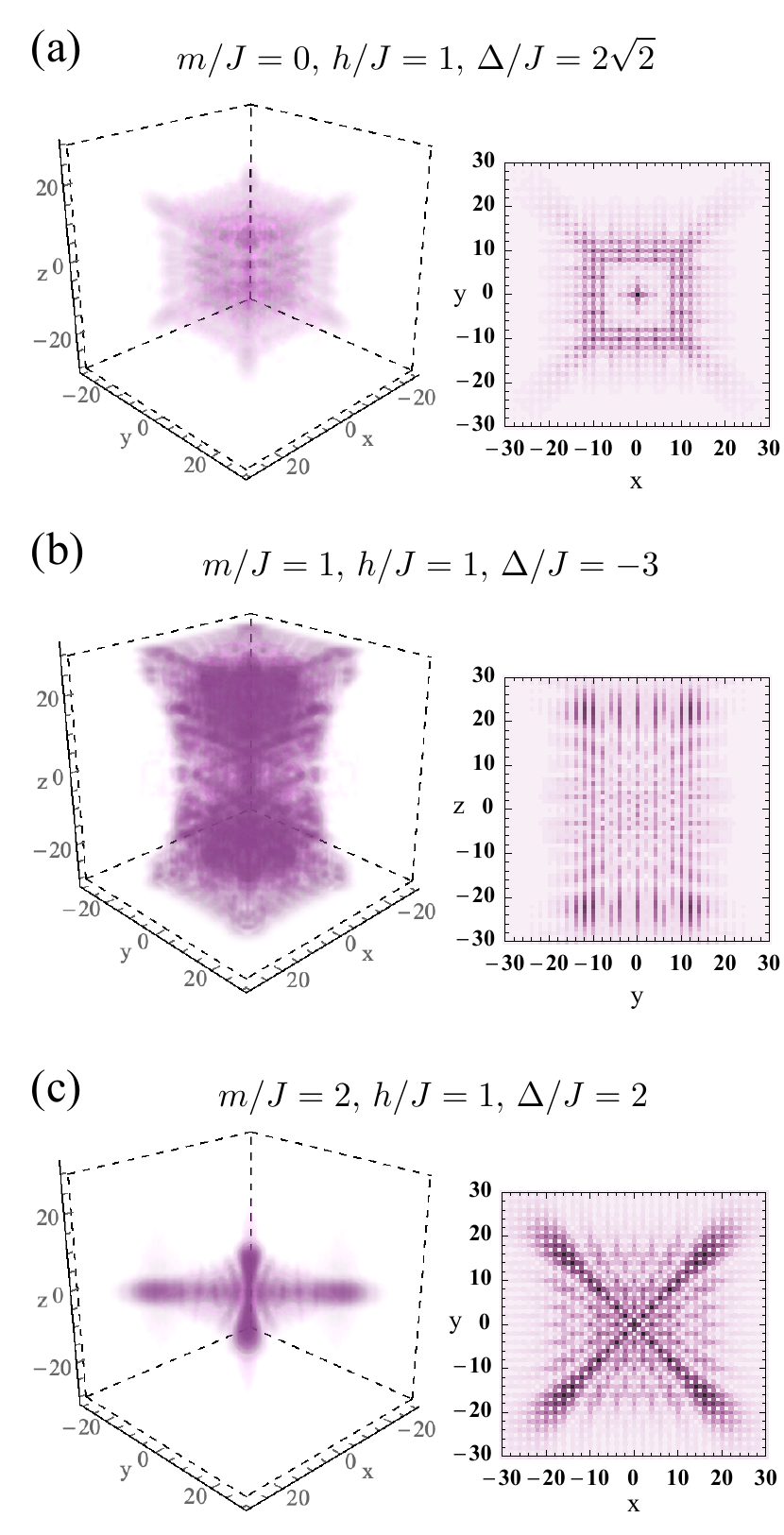}
	\caption{
	   Emission pattern at fixed time ($t/J=25$) of an emitter couple to a van Hove singularity for three different configurations of the system. For visualization, insets show special cuts along some selected directions: (a, c) $z=0$ and (b) $x=0$.
	   \vspace{-0.3cm}}
	\label{fig:Fig4}
\end{figure}

Main panels of Fig.~\ref{fig:Fig4} show the distribution of the photonic excitation a time $t/J=25$ after the emission has started, so that the emission pattern is clearly recognizable but no finte-size effects are present. Inset panels display specific cuts along some interesting planes. Figure.~\ref{fig:Fig4}(a) corresponds to the case where the staggered mass and hopping terms are given by $\stgmas=0$ and $\stghop/J=1$ respectively, and the emitter is detuned to the singularity at $\Delta/J=2\sqrt{2}$ (see red arrow in Fig.~\ref{fig:Fig3}(b)). In that case, light propagates uniformly in all Cartesian directions leading to cubic shaped emission pattern. Fig.~\ref{fig:Fig4}(b) accounts for the configuration in which $\stgmas/J=1$ and $\stghop/J=1$ respectively, and the emitter is detuned to the singularity at $\Delta/J=-3$ (see red arrow in Fig.~\ref{fig:Fig3}(d)), there we show the emission is specially intense along the vertical direction. Finally, Fig.~\ref{fig:Fig4}(c) displays the case corresponding to staggered mass and hopping terms given by $\stgmas/J=2$ and $\stghop/J=1$ respectively, and the emitter detuned to $\Delta/J=2$ (see red arrow in Fig.~\ref{fig:Fig3}(f)). In this case we obtain a highly anisotropic emission which is concentrated within the $z=0$ plane and that is specially intense along the $y=\pm x$ directions. These patterns show the tunability of the emission in these  photonic environments just by considering different emitter's detuning values.

\begin{figure*}[t]
	\centering
	\includegraphics[width=\textwidth]{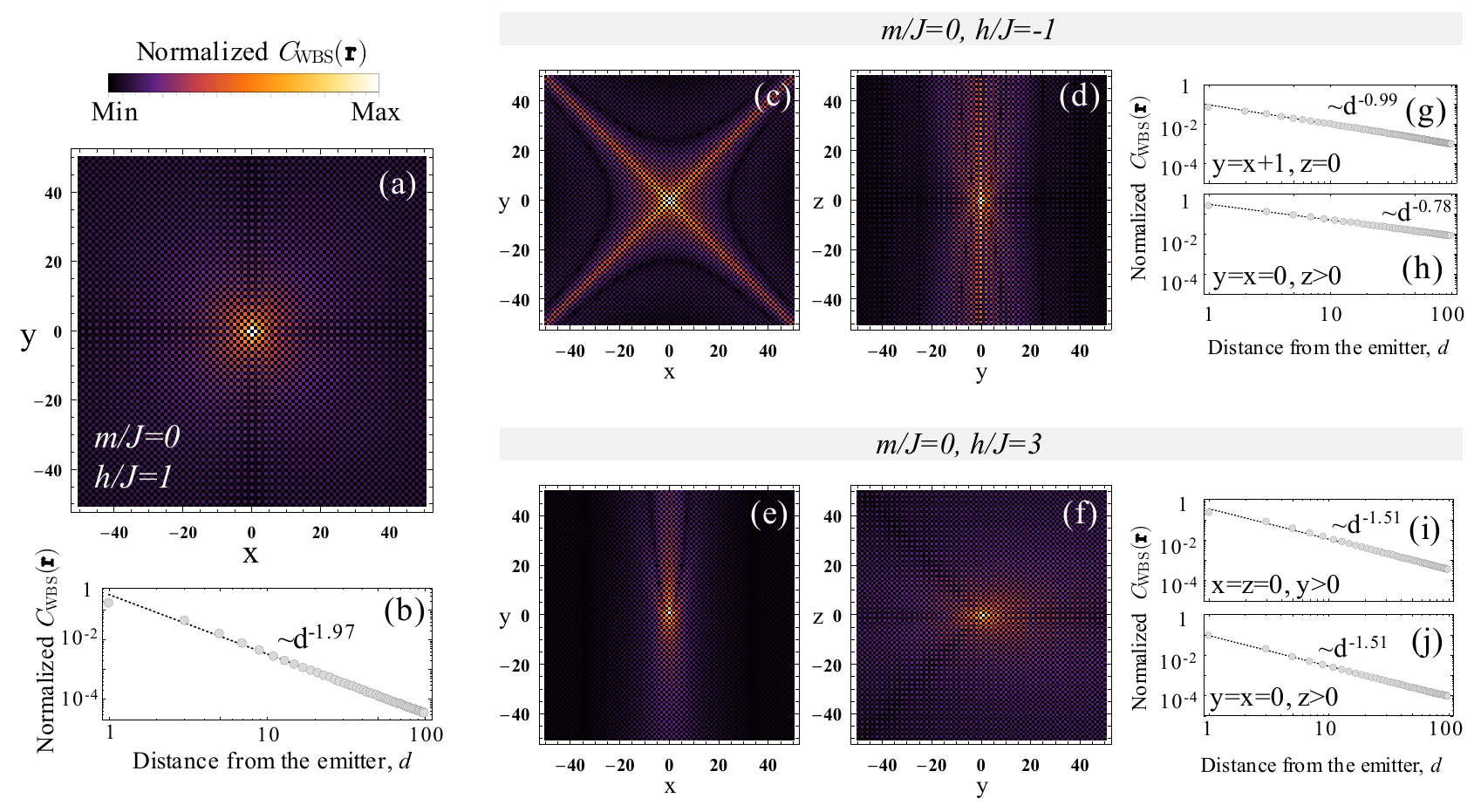}
	\caption{
	  (a) Cut along the $z=0$ plane of the photonic component associated to the Weyl bound state for the case $\stgmas=0$ and $\stghop/J=1$. Note that in this case the Weyl bound state confinement is mostly isotropic (b) Power law dependence of the bound state profile along any of the principal directions, i.e. $(\pm1,0,0)$,$(0,\pm1,0)$ or $(0,0,\pm1)$, for the case where $\stgmas=0$ and $\stghop/J=1$. (c-f) Cuts along the $z=0$ and $x=0$ planes of the photonic component associated to the Weyl bound state for the case $\stgmas=0$ and $\stghop/J=-1$ (c,d) or $\stghop/J=3$ (e,f). (g-h) Power law dependence of the Weyl bound state profile along some selected directions for the case where $\stgmas=0$ and $\stghop/J=-1$ (g,h) or $\stghop/J=3$ (i,j).
	   \vspace{-0.3cm}}
	\label{fig:Fig5}
\end{figure*}

\subsection{Bound states}

As we explained in Figs.~\ref{fig:Fig2}-\ref{fig:Fig3}, when the emitter's energy lies around the Weyl frequency its evolution displays a fractional decay. As aforementioned, this phenomena can be associated with the emergence of a localized photonic mode around the emitter that we label as Weyl bound-state~\cite{garcia-elcano2020a}. Mathematically, these bound states are connected to the existence of a real pole in $G_{e}(z)$ at a given energy $\omega$ (measured with respect to the Weyl frequency) that can be found by solving the following system of equations:
\begin{equation}
    \begin{cases}
    \omega-\Delta-\text{Re}\,\Sigma_{e}(\omega+i0^{+})=0\\
    \text{Im}\,\Sigma_{e}(\omega+i0^{+})=0
    \end{cases}
    \label{eq:RealPolesEquations}
\end{equation}

For the case  $\stgmas=0$ and $\stghop/J=1$, this equation is fullfilled for $\omega=0$ when the emitter's transition frequency coincides with the Weyl frequency ($\Delta=0$)~\cite{garcia-elcano2020a}. However the appearance of such Weyl bound state is not an exclusive feature of this particular system's configuration. In fact, inspection of the imaginary part of the single emitter self-energy for different staggered mass and hopping values shows that $\text{Im}\,\Sigma_e(\omega+i0^{+})$ goes to zero as $\omega \rightarrow 0$ irrespective of the selected parameters and the sublattice to which the emitter is coupled to. Then, given Eq.~\eqref{eq:RealPolesEquations}, one can guarantee the presence of a pole at the origin of the complex plane (that translates into the formation of the Weyl bound state) by setting the emitter's detuning to $\Delta=\text{Re}\,\Sigma(0+i0^{+})\equiv\Delta_c$ which we refer to as the critical detuning value. Note that, since $\Sigma_e(z)\propto\g^2/J$, the critical detuning remains close to zero if we restrict ourselves to the weak coupling regime, $g\ll J$.

Beyond the dynamical consequences for the single emitter dynamics, these bound states are relevant because they will mediate coherent exchange of excitations when many emitter's are coupled to the bath~\cite{douglas2015quantum,gonzalez2015subwavelength}. Thus, their spatial form is what will eventually determine the shape of the emitter-emitter interactions. In order to obtain the real space distribution of the photonic component of the Weyl bound state we can examine the eigenstates of the system whose associated eigenvalue coincides with the Weyl frequency. For that, we employ the time-independent Schr\"odinger equation: $H\ket{\psi_{\text{WBS}}}=E_{\text{WBS}}\ket{\psi_{\text{WBS}}}$, where $E_{\text{WBS}}$ must be taken to be the Weyl frequency. The projection of the photonic component of the Weyl bound state over a localize bosonic site is given by $C_{\text{WBS}}(\rr)=\bra{g;\mathrm{vac}}a_{\rr}\ket{\psi_{\text{WBS}}}$. In the following we will be using a normalize version of this quantity, with the normalization factor: $|C_e(t\rightarrow\infty)|(\g/J)$.

Figure~\ref{fig:Fig5} shows several real space profiles of the Weyl bound state corresponding to different choices of the staggered mass and hopping terms. 
Importantly, in contrast to the exponential localization featured by standard photon-atom bound states emerging in fully gapped systems~\cite{douglas2015quantum,gonzalez2015subwavelength}, the spatial confinement of the Weyl bound state follows a power-law dependence.
Although this power-law confinement also appears in other singular band-gap structures~\cite{gonzalez2018exotic,gonzalez2018non,perczel20a},  the topological protection of the Weyl degeneracies allows one to tune the power-law exponent~\cite{garcia-elcano2020a}. For instance, when the staggered mass and hopping terms are set to be $\stgmas=0$ and $\stghop/J=1$ respectively, the photonic component of the bound state displays an isotropic distribution around the emitter's position characterized by an inverse square power law as shown in Fig.~\ref{fig:Fig5}(a,b). As shown in~\cite{garcia-elcano2020a}, such isotropic behavior is lost for increasing values of $\stgmas$ leading to a strong confinement of the excitation along the vertical direction. In this work, we additionally demonstrate that one can further tune the shape of the Weyl bound state, without losing the power-law confinement, through the staggered hopping term $\stghop$. This is illustrated in Figs.~\ref{fig:Fig5}(c,d,e,f) that display the Weyl bound state's photonic component for configurations wherein no energy off-set between sublattices is considered but, instead, the staggered hopping term takes the non trivial values $\stghop/J=-1$ and $\stghop/J=3$. In particular, they show the $z=0$ and $x=0$ cuts of the three-dimensional distribution illustrating the main features characterizing these extreme cases. For $\stghop/J=-1$ we observe a highly directional confinement, specially along the diagonal directions in the $z=0$ plane and along the $z$ axis. More detailed inspection of the bound state profile along these directions reveals that the diagonal directions are characterized by an inverse linear power law whereas the vertical direction follows a power law featuring an even lower exponent (see Fig.~\ref{fig:Fig5}(g,h)). For $\stghop/J=3$ we observe that the bound state profile spreads over the $x=0$ plane, leading to a strong confinement along the $x$ direction. In this case, Fig.~\ref{fig:Fig5}(i,j) display the photonic component of the Weyl bound state as a function of the distance from the emitter along the positive $y$ and $z$ axis, which show the same power law dependence.

\section{Two emitters} \label{Two emitters}

Now, we investigate the scenario where two emitters are locally coupled to two different lattice sites in positions $\rr_1$ and $\rr_2$. 
The system is prepared such that only the emitter located at position $\rr_1$ is in its excited state whereas no photonic excitations are present: $\ket{\psi(0)}=\ket{e_1\,g_2;\mathrm{vac}}$.
Restricting ourselves to the single excitation subspace, we study the upper state population's dynamics of both the initially
excited emitter, $\left|U_{11}(t)\right|^2\equiv\left|C_1(t)\right|^2$, and the initially de-excited one, $\left|U_{21}(t)\right|^2\equiv\left|C_2(t)\right|^2$. The matrix elements of the problem's propagator can be obtained by solving the system of equations defined by Eq.~\eqref{eq:Propagator system of eqs} particularized for the case $\Nem=2$. They are given by:
\begin{equation}
\left(\begin{array}{cc}
G_{11}(z) & G_{12}(z)\\
G_{21}(z) & G_{22}(z)
\end{array}\right)
=\left(\begin{array}{cc}
z-\Delta_1-\Sigma_{11}(z) & -\Sigma_{12}(z)\\
-\Sigma_{21}(z) & z-\Delta_2-\Sigma_{22}(z)
\end{array}\right)^{-1}
\vspace{1mm}\\
\end{equation}
where $\Delta_{1/2}$ is the detuning of the initially excited/de-excited emitter with respect to the Weyl frequency and $\Sigma_{11}(z)$, $\Sigma_{12}(z)$, $\Sigma_{21}(z)$, and $\Sigma_{22}(z)$ can be calculated using Eq.~\eqref{eq:Self-energy matrix elements}.
Then, within the resolvent operator formalism, the dynamics of the initially excited and initially de-excited emitters can be computed through the inversion of  $G_{11}(z)\equiv G_{1}(z)$ and $G_{21}(z)\equiv G_{2}(z)$, respectively. Note that, if the initial condition had been $\ket{\psi(0)}=\ket{g_1\,e_2;\mathrm{vac}}$, the matrix elements $G_{12}(z)$ and $G_{22}(z)$ would play the role of $G_{1}(z)$ and $G_{2}(z)$ instead.

In the following we split the discussion in two parts: first we consider two emitters coupled to the same sublattice and then two emitters coupled to different sublattices. From now on, we assume that both the initially excited and the initially de-excited emitters have identical detuning values i.e. $\Delta_{1}=\Delta_{2}=\Delta$.

\subsection{Same sublattice AA/BB} \label{Same sublattice}

When the considered emitters are coupled to sites belonging to the same sublattice the initially excited/de-excited propagator can be decomposed as follows:
\begin{equation} \label{eq:G_{1/2}(z)}
    G_{1/2}(z) = \frac{1}{2}\left[G_{s}(z) \pm G_{a}(z)\right]
\end{equation}
where $G_{s/a}(z)=[z-\Delta-\Sigma_{s/a}(z)]^{-1}$ is the so-called symmetric/antisymmetric propagator and $\Sigma_{s/a}(z)$ is given by the sum/difference of the previously defined single emitter self-energy, $\Sigma_{e}(z)$, plus/minus an additional term, $\Sigma_{12}(z)$, which is sometimes referred to as the collective self-energy:
\begin{equation}
    \Sigma_{s/a}(z) = \Sigma_{e}(z)\pm\Sigma_{12}(z)\,.
\end{equation}
Comparing Eq.~\eqref{eq:G_{1/2}(z)} with the propagator obtained for the single emitter case (see Eq.~\eqref{eq:G_e(z)}) we realise that the dynamics of a pair of emitters --- provided that they are coupled to sites belonging to the same sublattice and feature identical detuning value --- consist in the linear combination of two independent single emitter problems. Therefore, the full machinery developed to tackle the single emitter case can be exploited. The only important modification that one has to perform is to include the collective self-energy contribution which, in the thermodynamic limit, reads:
\begin{equation} \label{eq:Sigma_1/2(z)}
    \Sigma_{12}^{(A/B)}(z) = 
    \frac{\g^{2}}{V_{\BZ}}\iiint_{\BZ}d^{3}k
    \frac{z \mp [\stgmas+2J\cos(k_z)]}{z^{2}-\omega(\kk)^2} e^{i\kk(\rr_2-\rr_1)}
\end{equation}
Note that the form of $\Sigma_{12}^{(A/B)}(z)$ resembles the one obtained for the single emitter self-energy (see Eq.~\eqref{eq:Single emitter Self-Energy (integral)}). Analogously, the superscript $A/B$ labels the sublattice to which the emitters are coupled to. However, the main difference with respect to the previously obtained single emitter self-energy, $\Sigma^{(A/B)}_e(z)$,  is the exponential factor introducing the dependence with the inter-emitter's distance. 

\subsubsection{Markovian limit}
As before, one can first study the dynamics within a Markovian description. For that we proceed as in the single emitter case, i.e. evaluating the self-energy's matrix elements at the emitters' detuning value: $\Sigma_{s/a}(\omega+i0^{+})\approx\Sigma_{s/a}(\Delta+i0^{+})$. After that, the inversion of the propagators $G_{1/2}(z)$ can be easily accomplished leading to the following expression for the $C_{1/2}(t)$ coefficient:
\begin{equation}
\begin{split}
    C_{1/2}(t) \approx &
    \;e^{-i\left[\Delta+\delta\omega_e(\Delta)-i\frac{\Gamma_e(\Delta)}{2}\right]} \\
    & \times \frac{1}{2}\left(
    e^{-i\left[\delta\omega_{12}(\Delta)-i\frac{\Gamma_{12}(\Delta)}{2}\right]}
    \pm
    e^{+i\left[\delta\omega_{12}(\Delta)-i\frac{\Gamma_{12}(\Delta)}{2}\right]}
    \right)
\end{split}
\end{equation}
where 
$\delta\omega_e(\Delta)=\mathrm{Re}\,\Sigma_e(\Delta)$ and $\delta\omega_{12}(\Delta)=\mathrm{Re}\,\Sigma_{12}(\Delta)$ stand for the single emitter and collective energy shifts, respectively, whereas 
$\Gamma_{e}(\Delta)=-2\,\mathrm{Im}\,\Sigma_{e}(\Delta)$ and
$\Gamma_{12}(\Delta)=-2\,\mathrm{Im}\,\Sigma_{12}(\Delta)$ denote the single emitter and collective decay rates, respectively. 
Inspection of the self-energy's matrix elements shows that both the single emitter and collective decay rates vanish when the emitters are detuned to the Weyl frequency. Therefore, for this particular detuning value, the Markovian treatment of the problem anticipates an oscillatory behavior revealing a continuous exchange of the excitation between the two emitters:
\begin{equation}
    \left|C_{1/2}(t)\right|^2 \approx \cos^2(J^{\mathrm{M}}_{12}t)\:\Big/\:\sin^2(J^{\mathrm{M}}_{12}t)
\end{equation}
Here, we have identified $J^{\mathrm{M}}_{12}=|\delta\omega_{12}(0)|$ as the exchange rate of the excitation between emitters within the Markovian limit.

\subsubsection{Non-Markovian dynamics}\label{2QEs-Same sublattice-Non Markovian}

As we know from the single emitter situation, the Markovian description can deviate significantly from the exact dynamics in certain parameter regimes. In order then to assure this remarkable coherent exchange of excitations with no associated dissipation, we will perform exact calculations using the resolvent operator method. For that, we focus on a system's configuration wherein the quantum emitters are assumed to be coupled to two vertically aligned sites belonging to the same sublattice, and where the staggered hopping term is $\stghop/J=1$. The distance between emitters is then given by the $z$-component of the vector that connects them, $z_{12}$. Within this configuration, we can integrate the $k_x$ and $k_y$ directions of the collective self-energy (Eq.~\eqref{eq:Sigma_1/2(z)}) obtaining:
\begin{widetext}
\begin{equation}
    \Sigma_{12}^{(A/B)}(z) =
    \frac{2\g^2}{\pi^2}
    \int_{0}^{\pi}dk_{z}
    \frac{z\mp [\stgmas+2J\cos(k_z)]}{z^2-4J^2-[\stgmas+2J\cos(k_z)]^2}
    K\left[\left(\frac{4J^2}{z^2-4J^2-[\stgmas+2J\cos(k_z)]^2}\right)^2\right]\cos(k_z z_{12})
\end{equation}
\end{widetext}
%

\begin{figure}[tb]
	\centering
	\includegraphics[width=\columnwidth]{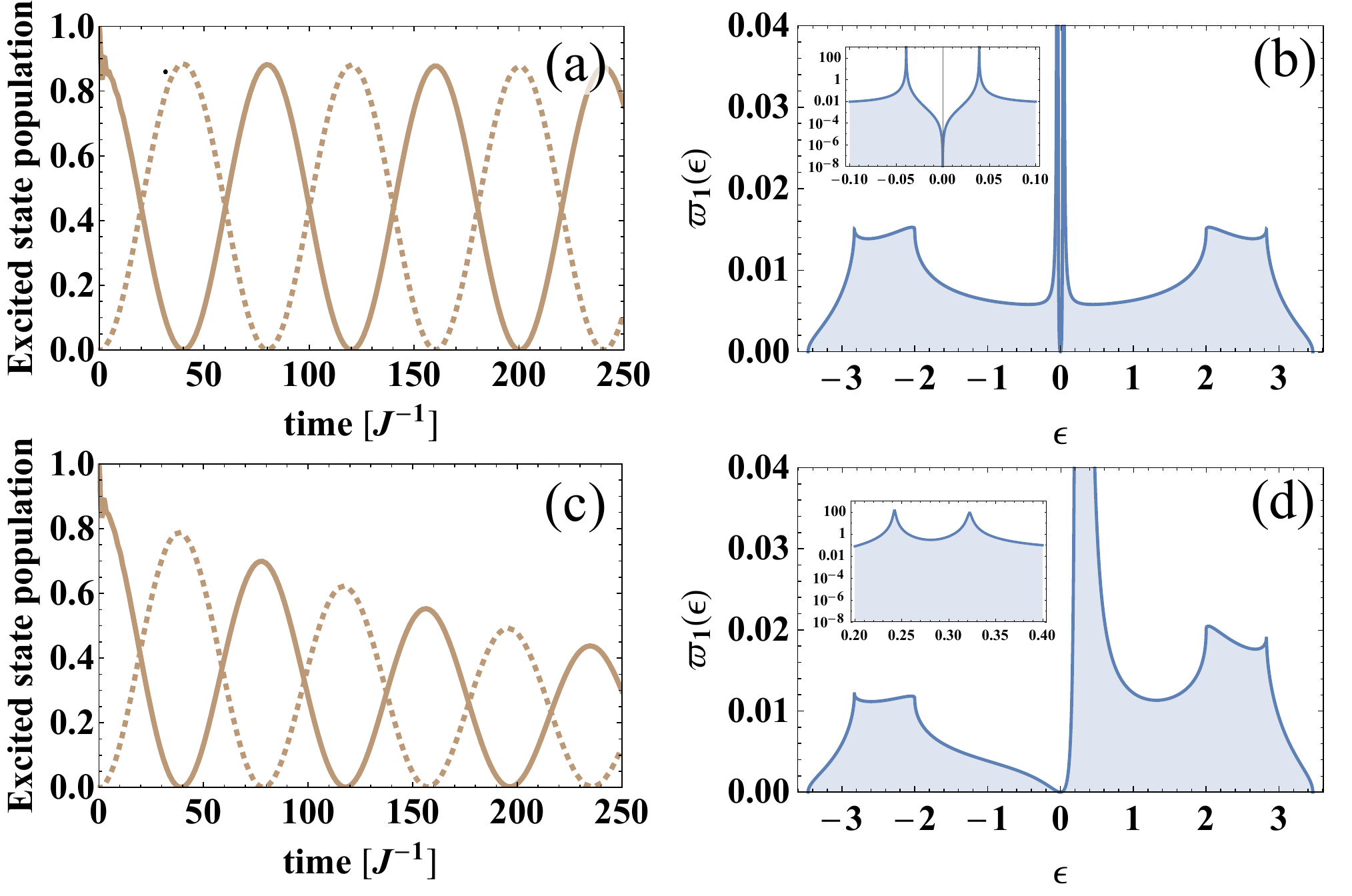}
	\caption{
	    Dynamical behavior of the two emitter problem for the case where no staggered mass nor staggered hopping terms are considered i.e. $\stgmas=0$ and $\stghop/J=1$. The light matter coupling is chosen to be $\g/J=0.5$
	    (a) Temporal evolution of the initially excited emitter (solid line) and of the initially de-excited one (dashed line) for $\Delta/J=0$.
	    (b) Spectral density function of the initially excited emitter associated to the dynamics shown in panel (a).
	    (c) Temporal evolution of the initially excited emitter (solid line) and of the initially de-excited one (dashed line) for $\Delta/J=3$.
	    (d) Spectral density function of the initially excited emitter associated to the dynamics shown in panel (c).
	   \vspace{-0.6cm}}
	\label{fig:Fig6}
\end{figure}

With this expression at hand we can calculate the temporal evolution of a pair the initially excited (straight line) and initially de-excited (dotted line) emitters for two different detuning values: $\Delta/J=0$ and $\Delta/J=0.3$, respectively, shown in panels (a) and (c) of Fig.~\ref{fig:Fig6}. Here we have assumed that both emitters couple to the A sublattice, although similar behaviour occurs if they couple to the B sublattice. For both detuning values we observe that the initially excited emitter undergoes a fast relaxation at very short times followed by a set of oscillations. For larger times important differences are found between the two displayed cases. On one hand, when the emitters are detuned to the Weyl frequency, the amplitude of the observed oscillations remains unchanged. On the other hand, when the emitters are detuned inside the band we observe a clear attenuation of the oscillations after three complete cycles. Note that these calculations are made for fixed values of the light-matter coupling ($\g/J=0.5$), the staggered mass and hopping terms ($\stgmas/J=0$ and $\stghop/J=1$), and the vertical separation between emitters ($z_{12}=1$).

Disregarding the short time behavior of the system, the temporal evolution of the initially excited/de-excited emitter is reliably reproduced by the following expression:
\begin{equation}
    |C_{1/2}(t)|^2 \approx \left.
    R\,e^{-\gamma_{12}t}\,\cos^2(J_{12}\:t)
    \:\Big/\: 
    R\,e^{-\gamma_{12}t}\,\sin^2(J_{12}\: t)\right.
\end{equation}
where $R$ coincides with the fractional steady-state population value which characterizes the emergence of the Weyl bound state in the single emitter case. This observation is of paramount importance since it indicates that the physical mechanism which mediates the excitation's exchange is the emergence of the Weyl bound states \cite{garcia-elcano2020a}. Besides, we refer to $\gamma_{12}$ and $J_{12}$ as the dissipative and coherent components of the two emitters' dynamics. In order to have a complete oscillatory behavior it is necessary that $\gamma_{12} \ll J_{12}$, as it is the case for the two considered detuning values. However, we also have that $0<\gamma_{12}(\Delta=0) \ll \gamma_{12}(\Delta=0.3)$, that is, the dissipative component increases as we detune the emitters away from zero energy. Importantly, using an exact treatment we observe that even when the emitters are in resonance with the Weyl frequency the dissipative term does not completely vanish, which is in direct contradiction with the Markovian prediction. Intuitively, this can be understood from the fact that the interaction between the emitters through the bath modes induces a displacement of the emitters' energies from the Weyl point, thus introducing a small dissipative component in the dynamics.
To clarify this point, in the following, we study the ratio between the dissipative and the coherent component of the dynamics using the alternative calculation of the dynamics based on the spectral density function, i.e. $\varpi_{1/2}(\omega)=-\frac{1}{\pi}\mathrm{Im}\,G_{1/2}(\omega+i0^{+})$.

\begin{figure}[tb]
	\centering
	\includegraphics[width=\columnwidth]{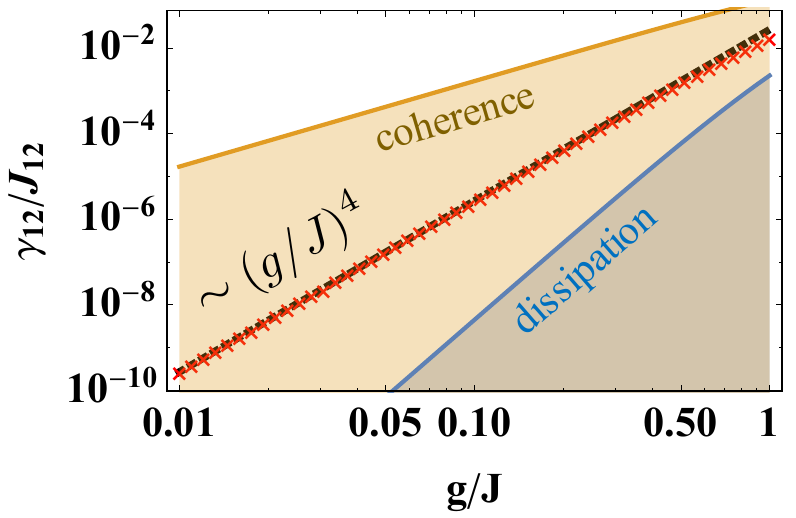}
	\caption{
	   Red crosses denote the ratio between the dissipative component of the dynamics, $\gamma_{12}$, and the coherent component, $J_{12}$, for the considered range of light-matter coupling values. Black dotted line stands for the corresponding linear fitting which yields a $\sim(\g/J)^4$ dependence. As shown the coherent component (yellow shaded area) is always much larger that the dissipative component (blue shaded area).
	   \vspace{-0.5cm}}
	\label{fig:Fig7}
\end{figure}

As we show in appendix~\ref{appendix:inversion}, the temporal evolution associated to the initially excited/de-excited emitters can be calculated by Fourier transforming the corresponding spectral density function. Thus, we can gain some intuition of the studied problem exploring these quantities. In particular, panels (b) and (d) of Fig.~\ref{fig:Fig6} depict the spectral density functions of the initially excited emitter, $\varpi_{1}(\omega)$, for the two considered detuning values and the selected system configuration. As required by the initial condition, $\varpi_{1}(\omega)$ is normalized to one. Also, we observed that this quantity goes to zero as $\omega\rightarrow 0$. But the most remarkable feature of $\varpi_{1}(\omega)$ is that it presents two peaks whose mathematical origin is related to the poles of the symmetric and anti-symmetric propagators:
\begin{equation}
    z-\Delta-\left[\Sigma_e(z)\pm\Sigma_{12}(z)\right]=0
\end{equation}
Interestingly, we can naturally connect the exchange rate of the observed oscillations to the spectral separation between the peaks and the dissipation component of the dynamics to the width of the peaks. Taking that into account we propose a model for the spectral density function in the proximities of the Weyl frequency ($|\omega|\ll J$) which reads:
\begin{equation}
    \varpi_{1}(\omega)\approx\sum_{\alpha=s,a}D_{\alpha}\frac{(w_\alpha/2)^2}{(\omega-\omega_{\alpha})^2+(w_\alpha/2)^2}\left(\frac{\omega}{\omega_\alpha}\right)^q
\end{equation}
where $w_{s/a}$ and $\omega_{s/a}$ control the width and the position of the peaks, respectively, and $D_{a/s}$ is a normalization constant. The last factor reproduces the dependence of the density of states of the bath around the Weyl frequency, e.g. for $\stgmas=0$ we must set $q=2$. It must be noticed that the dissipative and coherent components of the dynamics can be calculated as follows: $\gamma_{12}=\frac{1}{2}(w_s+w_a)$ and $J_{12}=\frac{1}{2}(\omega_s-\omega_a)$. Hence, comparing the proposed model to the spectral density functions computed from the propagator associated to the initially excited emitter we obtain the specific values of the dissipative and coherent components of the dynamics for any given configuration. Finally, in Fig.~\ref{fig:Fig7} we plot the ratio $\gamma_{12}/J_{12}$ versus the light-matter coupling. We observe that the dissipative component always features a much smaller value that its coherent counterpart. For this particular system configuration we have that $\gamma_{12}/J_{12} \sim (g/J)^4$, which therefore tell us that the Markovian approximation will be valid as long as $(g/J)^4\ll 1$ is satisfied.

\subsection{Different sublattice AB} \label{Different sublattice}

\begin{figure}[tb]
	\centering
	\includegraphics[width=8cm]{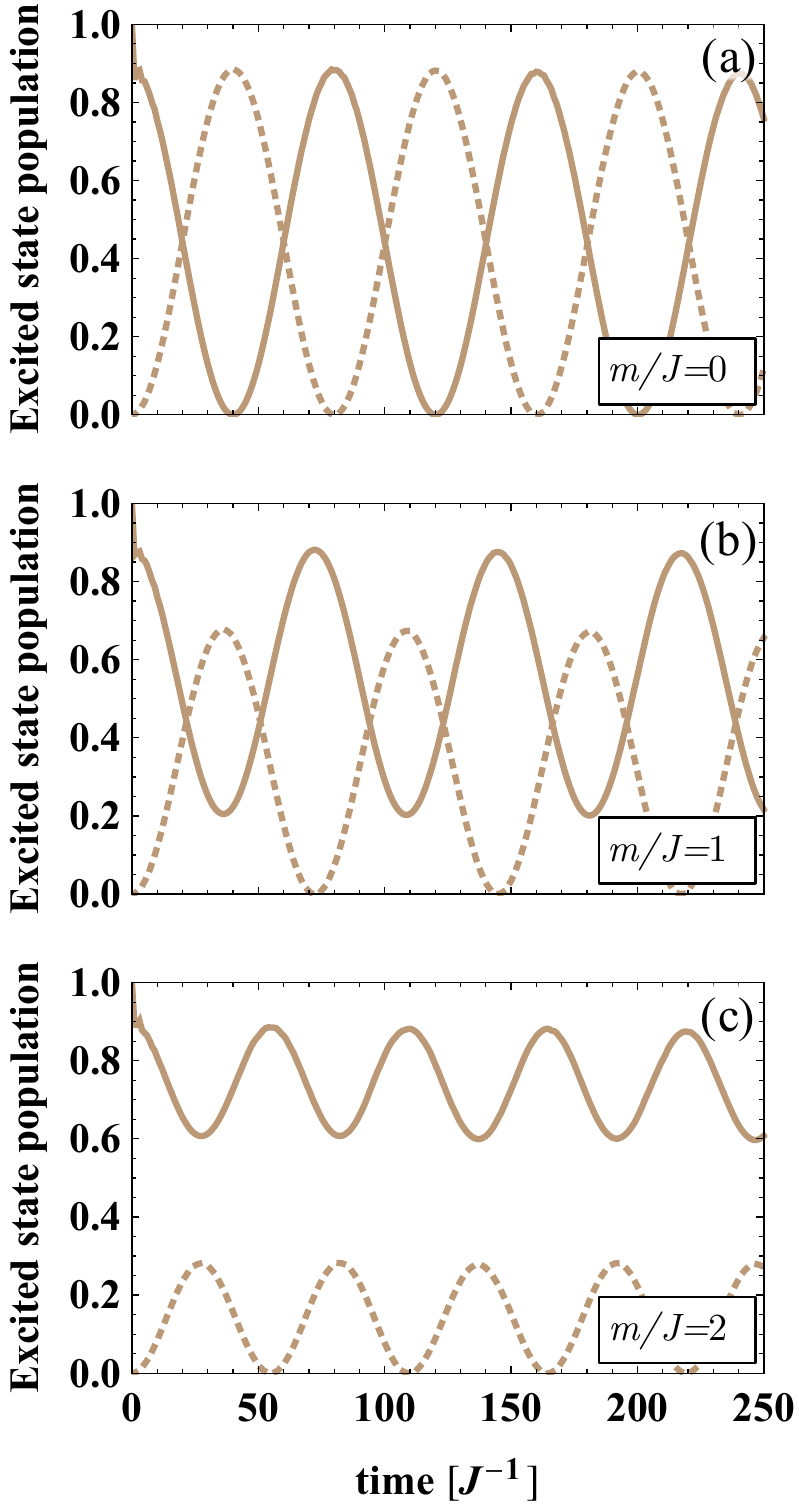}
	\caption{
	    (a) Dynamics of the initially excited (solid line) and the initially de-excited emitter (dashed line) for a system configuration in which the emitters are coupled to two sites belonging to a different sublattice separated a distance equal to the lattice parameter along the $x$ direction. Both emitters are detuned towards the Weyl frequency and the parameters of the bath are chosen to be $\stgmas=0$ and $\stghop/J=1$.
	    (b) The same as in (a) but, in this case, the staggered mass term is chosen to be $\stgmas/J=1$.
	    (c) The same as in (a) but, in this case, the staggered mass term is chosen to be $\stgmas/J=2$.
	    \vspace{-0.5cm}
	    }
	\label{fig:Fig8}
\end{figure}

The decomposition of the propagator's matrix elements accounting for the dynamics of the initially excited and the initially de-excited emitters given by Eq.~\eqref{eq:G_{1/2}(z)} is no longer valid when the emitters are coupled to sites belonging to different sublattices. This condition complicates substantially the quest for solutions based on the resolvent operator formalism. However, numerical strategies can be readily implemented leading to equivalent results. 
The dynamical behavior associated to two quantum emitters couple to sites belonging to distinct sublattices will, in general, reproduce the same phenomenology observed for the case in which the emitters are coupled to sites belonging of the same sublattice, that is, a continuous exchange of excitations mediated by the Weyl bound states around the emitters. This is indeed what we observe in the example chosen of Fig.~\ref{fig:Fig8}(a)  when the staggered mass term is set to zero. However, when $\stgmas\neq 0$ the behaviour changes dramatically, as shown in Figs.~\ref{fig:Fig8}(b-c), where we see that the initially excited state does not ever get completely de-excited whereas the initially de-excited emitter does not ever reach the maximal value of the oscillations which is established by the stationary state population that is found within the single emitter analysis. The underlying reason is that when $\stgmas\neq 0$, the individual energy shifts $\delta\omega_e^{A/B}(\Delta)$ appearing because of the coupling to the bath are different for the emitters coupled to the A and B sublattices. This creates an effective detuning between emitters,  with initially the same energy, which leads to incomplete coherent exchange oscillations. This is a relevant finding that will also affect to other reservoirs where such asymmetric energy shifts occur~\cite{gonzalez2018anisotropic}.

\section{Many emitters: effective spin model description} \label{Many emitters}

Analyzing the exact dynamics in the many emitter configurations can be done in an exact way following similar procedures as the one we use in the previous section. Instead of considering some particular situations to illustrate it, here we will rather present how the system will be described within a Markovian treatment, and generalize the arguments given for the two-emitter case to obtain its regime of validity. As shown in Refs.~\cite{douglas2015quantum,gonzalez2015subwavelength}, an alternative way of describing the effect of the bound-state mediated interactions consists in adiabatically eliminating the bath degrees-of freedom under the Born-Markov approximation. Doing this adiabatic elimination for our bath results into an effective spin-model description without any associated dissipative term which reads:
\begin{align}
 H_{\mathrm{spin}}=\sum_{j.\alpha} \delta\omega_e^\alpha\sigma_{ee}^j +\sum_{jj^{\prime}} J_{jj^{\prime}}^{\alpha\alpha^{\prime}}\,\sigma_{eg}^{j} \, \sigma^{j^{\prime}}_{ge}\,,
 \end{align}
 where $J_{jj^{\prime}}^{\alpha\alpha^{\prime}}$ represents the interaction between the pair of emitters $j$ and $j^{\prime}$ ($\alpha,\alpha^{\prime}=A/B$ denotes the sublattice to which the corresponding emitter belongs to), and $\delta\omega_e^\alpha$ is the sublattice-dependent energy-shift of the $j$-th emitter. The crucial point to realize is that, in the studied system, $J_{jj^{\prime}}^{\alpha\alpha^{\prime}}$ inherits the dependence with the distance between the two emitters ($|\mathbf{r}_j-\mathbf{r}_{j^\prime}|$) from the space dependence of the bound-state wavefunction, i.e., $J_{jj^{\prime}}^{AA(AB)}\propto C_{\mathrm{WBS}}(\mathbf{r}_j-\mathbf{r}_{j^\prime})$. Note that, consistently, our model captures both the complete and incomplete coherent oscillations observed in the previous section. The regime of validity of these description will depend on the particular situation chosen, however, we can try to estimate them generalizing the argumentation's done for the two-emitter situation. When $\Nem$ emitter are present, the maximum dissipation that can appear in the single-excitation scales with a collective factor $\sim \Nem \gamma_{12}$, where $\gamma_{12}$ is the largest pairwise dissipation appearing in a given configuration, as defined in section~\ref{2QEs-Same sublattice-Non Markovian}. Then, e.g. for the $\stgmas=0$ and $\stghop/J=1$ case, we would have that $\gamma_{12}^{\text{col}}/J_{12}^{\text{col}} \sim \Nem(g/J)^4$. Therefore, considering the limit $g/J \ll 1$ will still yield a proper framework in which the dissipative component of the dynamics can be chosen to be neglected with respect to the corresponding coherent counterpart.

\section{Conclusions and outlook} \label{Conclusions}

Summing up, we have studied extensively the quantum optical phenomena that emerges when emitters couple to the bulk modes of a photonic Weyl environment. By doing that, we have found several important results, such as the possibility of tuning the Weyl bound states (and emitter interactions) through the consideration of both a staggered mass and a staggered hopping term in the bath's Hamiltonian. We also unveiled the appearance of an asymmetric dynamical behaviour for the upper/lower bands in the single emitter case, accompanied by directional emission patterns, as well as the emergence of incomplete exchange oscillations when two emitter couple to different sublattices. Beyond that, we also derived an effective spin model description when many emitters are coupled simultaneously to the bath and tuned to the Weyl point frequency. We believe our work opens up several interesting research directions. One option consists in studying the many-body phases of the interacting spin models obtained with these topological baths, as recently explored in 1D settings~\cite{bello2019a}. Besides, one can search for novel super/subradiant phenomena~\cite{Gonzalez-Tudela2017} that can appear because of the directional radiation patterns appearing when the emitters are tuned to special points of the band-structure.

\section*{Acknowledgments.}
AGT   acknowledges   support   from   CSIC Research   Platform   on   Quantum   Technologies   PTI-001  and  from  Spanish  project  PGC2018-094792-B-100(MCIU/AEI/FEDER, EU). J.B.-A. and I.G.-E. acknowledge financial support from the Spanish Ministry for Science, Innovation, and Universities through grants RTI2018-098452-B-I00 (MCIU/AEI/FEDER,UE), the “Mar\'ia de Maeztu” programme for Units of Excellence in R\&D (MDM-2014-0377) and FPU grant AP-2018-02748.

\begin{appendices}

\section{Inversion of $G_{jj^{\prime}(z)}$} \label{appendix:inversion}

\begin{figure}[tb]
	\centering
	\includegraphics[width=8cm]{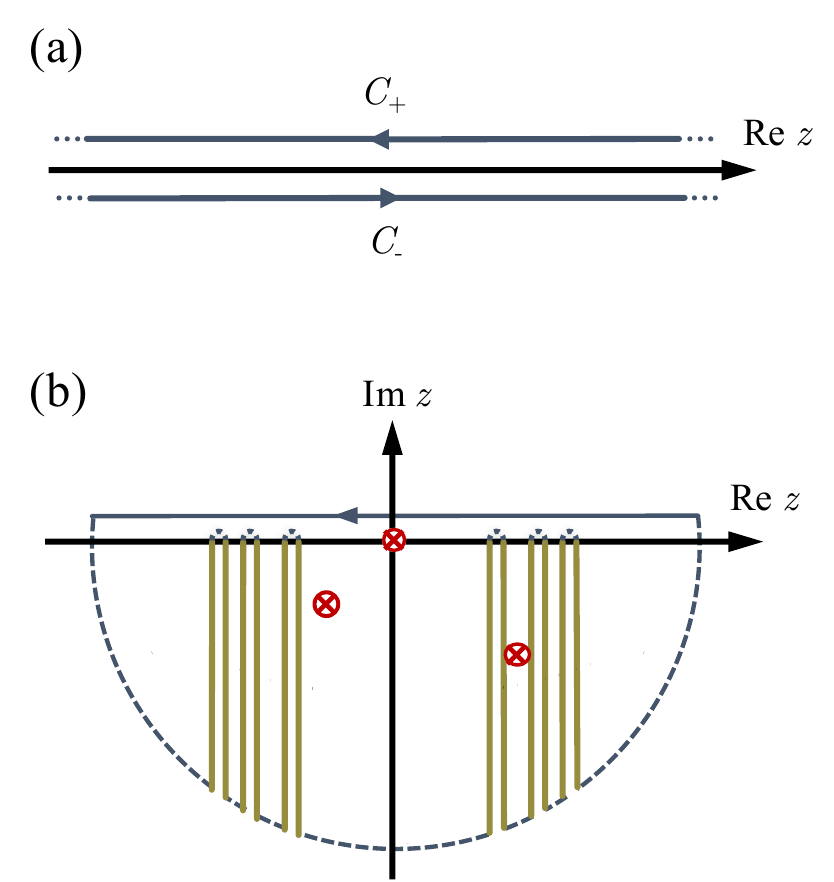}
	\caption{
	   (a) Retarded ($C_+$) and advanced ($C_-$) contributions to the dynamics. 
	   (b) Outline of the integration path employed to compute the dynamics of the system for the configuration where $\stgmas=0$ and $\stghop/J=1$. Solid brown lines stand for the detours that must be performed to avoid the non-analytical regions of the problem's self-energy. Red crosses indicate the potential positions of the poles featured by the studied propagator.
	   \vspace{-0.5cm}}
	\label{fig:Fig1A}
\end{figure}

The inversion given by Eq.~\eqref{eq:Propagator inversion} can be accomplished employing the residue theorem. For that, assuming $t>0$ so that the advanced contribution is zero, we introduce a complex path including a straight line above the real axis whose extremes are connected by a semicircle enclosing the lower half plane. As shown in Fig~\ref{fig:Fig1A}(b) we must also deform the contour to avoid the non analytical regions of the propagator which emerge due to the presence of van-Hove singularities in the bath spectrum.
Namely, the transition probability $U_{jj^{\prime}}(t)$ can be written formally expressed as:
\begin{equation}\label{eq:Poles and Detour contributions}
    U_{jj^{\prime}}(t)=I_{\mathrm{Poles}}(t)+I_{\mathrm{Detour}}(t)
\end{equation}
where, $I_{\mathrm{Poles}}(t)$ is explicitly given by:
\begin{equation}
    I_{\mathrm{Poles}}(t)=\sum_{p}R_{p}\,e^{-iz_{p}t}
\end{equation}
with
\begin{equation}
    R_{p}=\frac{1}{1-\left.\frac{\partial}{\partial z} \Sigma_{jj^{\prime}}\left(z\right)\right|_{z=z_{p}}}
\end{equation}
such that $p$ runs over the number of poles with position $z_{p}$ and residue $R_{p}$. This component is traditionally divided in two contributions, $I_\mathrm{Poles}=I_\mathrm{BS}(t)+I_\mathrm{UP}(t)$, depending on whether the existing poles are purely real (bound states) or whether they feature a finite imaginary part (unstable poles). On the other hand, the detour contribution is given by:
\begin{equation}
\begin{split}
    & I_{\mathrm{Detour}}(t)= \\
    & \quad\sum_{n}\frac{e^{-ix_{n}t}}{2\pi}\int_{0}^{\infty}dy\left[G_{jj^{\prime}}\left(x_{n}^{+}-iy\right)-G_{jj^{\prime}}\left(x_{n}^{-}-iy\right)\right]e^{-yt}
\end{split}
\end{equation}
where $n$ runs over the number of van Hove singularities and $x_{n}^{+/-}$ denotes the position in the real axis of such singularities plus/minus an infinitesimal quantity: $x_{n}^{\pm}=\left[x_{n}\pm\epsilon\right]_{\epsilon\rightarrow0}$.
Remarkably, this strategy provides an elegant physical interpretation of the undergoing phenomena but requires a thorough knowledge of the complex structure of the problem's self-energy. 

Noteworthy, the same results can be obtained by Fourier transforming the matrix elements of the so-called spectral density function, defined as:
\begin{equation}
\varpi_{jj^{\prime}}(\omega)\equiv-\frac{1}{\pi}\text{Im}\,G_{jj^{\prime}}(\omega+i0^{+})\,,
\end{equation}
providing us an alternative way of calculating the dynamics without the use of residue theorem, but still within the framework of the resolvent method. This representation can be directly derived from Eq.~\eqref{eq:Propagator inversion} if we formally retain both the $C_{+}$ and the $C_{-}$ contributions and realize that $G_{jj^{\prime}}(\omega-0^{+})-G_{jj^{\prime}}(\omega+0^{+})=-2i\,\mathrm{Im}\left\{G_{jj^{\prime}}(\omega+0^{+})\right\}$. In the last equality we have used that $G_{jj^{\prime}}(z^{*})=G_{jj^{\prime}}^{*}(z)$ which is a general property of the resolvent ensured by the Hermiticity of $H_{0}+H_{I}$.

\section{Numerical methods} \label{appendix:numerics}

As stated in the main text, an alternative strategy to compute the emitters' dynamics consists in solving the Schr\"{o}dinger equation for the total Hamiltonian: $H=H_M+H_B+H_I$. There are several methods to do that efficiently. One approach consists in using split-methods that perform the evolution of the bath/emitter Hamiltonian in reciprocal/real space sequentially~\cite{Gonzalez-Tudela2017a}. In our case, we will use an approach that performs the full evolution in real space, such that the first step is defining an appropriate matrix description of the full Hamiltonian $\mathbb{H}$. For that, we use a basis of $N_B^3$ localized bosonic modes and restrict ourselves to the single excitation subspace, such that $H_B$ can be represented as a square matrix of dimension $N_B^3 \times N_B^3$. Since the interaction is assumed to be local, we can incorporate the emitter degrees of freedom straightforwardly and connect them to specific bosonic sites. The final result is that the time-dependent Schr\"{o}dinger equation for the considered problem can be formulated as a system of $N_B^3+\Nem$ differential equations:
\begin{equation}\label{eq:System of ODEs}
    \dot{\bm{\psi}}(t)=-i \mathbb{H} \,\bm{\psi}(t)
\end{equation}
where $\bm{\psi}(t)\in\mathbb{C}^{N_B^3+\Nem}$ is a time-dependent vector defined as $\bm{\psi}(t)=(C_{\rr_1}(t),...,C_{\rr_{N_B^3}}(t),C_1(t),...,C_{\Nem}(t))^T$ with $C_{\rr_i}(t)=\bra{\Psi_0}\bd{\rr}\ket{\Psi(t)}$ and $C_j(t)$ defined as in the previous section. The matrix representing the total Hamiltonian of the system has the following structure:
\begin{equation}
\mathbb{H} =
\begin{tikzpicture}[baseline=(current bounding box.center)]
\matrix (m)[
    matrix of math nodes,
    left delimiter={(},right delimiter={)},
    inner sep=1pt,
    column sep=0.5pt,row sep=0.5pt
    ] 
    {|[inner sep=5mm]|H_B & |[inner sep=1mm]|H_I\\
     |[inner sep=1mm]|H_I & |[inner sep=1mm]|H_M\\};
     
\draw (m-1-1.south west) |- (m-1-1.north east) |- (m-1-1.south west);
\draw (m-2-2.south west) |- (m-2-2.north east) |- (m-2-2.south west);
\draw (m-2-2.north east) |- (m-1-1.north east);
\draw (m-1-1.south west) |- (m-2-2.south west);
\end{tikzpicture}
\end{equation}
where the block representing the bath Hamiltonian, $H_B$, is usually a sparse matrix, the block representing the matter Hamiltonian, $H_M$, is diagonal, with each entry accounting for the detuning of the corresponding emitter with respect to the Weyl frequency, and the blocks representing the interaction Hamiltonian have non zero entries at the lattice sites to which each of the considered emitters is coupled to. The solution to the system of differential equations given by Eq.\eqref{eq:System of ODEs} is given by the action of an exponential matrix over a given vector representing the initial condition:
\begin{equation}
        \bm{\psi}(t)=\exp\left(-i\,\mathbb{H}\,t\right)\,\bm{\psi}(0)\,,
\end{equation}
that can be efficiently computed using, for example, the algorithm developed in \cite{al-mohy2011a}.

\section{Multivalued structure of $\Sigma_e(z)$} \label{appendix:Multi-valuation}

In this section, we analyse the complex structure of self-energy matrix element given by Eq.~\eqref{eq:Sigma_e(M=0)} which corresponds to the case of a single emitter interacting with a bath wherein no staggered mass nor hopping terms are considered. First, we realise that $\Sigma_{e}(z)$ is a multivalued function so that, in order to obtain the associated dynamics using complex integration techniques, one needs to accomplish a convenient mapping of the complex energy plane into meaningful branches of the multivalued self-energy. The final goal is to define a single valued propagator which can be directly plugged in Eq.~\eqref{eq:Propagator inversion} leading to comprehensible results when the detour integration is performed. To do that let us start introducing a special notation to refer to the different branches of the complex elliptic integral:
\begin{equation}
K^{\left[p,q\right]}\left(z\right)=p\,K^{\left[0,0\right]}\left(z\right)+2q\,K^{\left[0,0\right]}\left(1-z\right)
\end{equation}
where $K^{\left[0,0\right]}(z)$ stands for the first Riemann sheet of the complex elliptic integral of the first kind. Then, to simplify further implications, let us introduce the following quantity:
\begin{equation}
\Sigma_{\alpha \beta}^{[q]}(z)=\frac{4\g^{2}}{\pi^{2}}\frac{z}{z^{2}-6J^{2}}
\frac{1-9\xi_{\alpha \beta}^{4}}{\left[1-\xi_{\alpha \beta}\right]\left[1+3\xi_{\alpha \beta}\right]}\,\left[K^{[1,q]}\left(m_{\alpha \beta}\right)\right]^{2}
\end{equation}
where we have defined:
\begin{equation}
m_{\alpha \beta}=\frac{16\xi_{\alpha \beta}^{3}}{(1-\xi_{\alpha \beta})(1+3\xi_{\alpha \beta})} 
\end{equation}
with
\begin{equation}
\xi_{\alpha\beta}=\frac
{\sqrt{1+\alpha\sqrt{1-\frac{1}{9}\left(\frac{6J^{2}}{z^{2}-6J^{2}}\right)^{2}}}}
{\sqrt{1+\beta\sqrt{1-\left(\frac{6J^{2}}{z^{2}-6J^{2}}\right)^{2}}}}
\end{equation}
Provided that, $\sqrt{\quad}$ denotes the first Riemann sheet of the complex square root, the defined quantity is specified by three labels $q\in\mathbb{Z}$ and $\alpha,\beta\in\left\{ -1,+1\right\}$. It must be noticed that, for a given combination of $q$, $\alpha$ and $\beta$, $\Sigma_{\alpha\beta}^{\left[q\right]}$ is a single valued function of the complex value z. Making use of the introduced notation, we conveniently defined six different Riemann branches of the multivalued self-energy as follows:
\begin{equation}
\begin{array}{ccl}
\Sigma_{e}^{I}(z)&=&\Sigma_{-+}^{\left[0\right]}\vspace{2mm}\\
\Sigma_{e}^{II}(z)&=&
\begin{cases}
\Sigma_{--}^{\left[0\right]} & x^{2}-y^{2}>6\\
\Sigma_{--}^{\left[1\right]} & x^{2}-y^{2}<6
\end{cases}\vspace{2mm}\\
\Sigma_{e}^{III}(z)&=&
\begin{cases}
\Sigma_{+-}^{\left[0\right]} & 1-6r^{2}+3r^{4}+2\cos\left(2\theta\right)<0\\
\Sigma_{+-}^{\left[-1\right]} & 1-6r^{2}+3r^{4}+2\cos\left(2\theta\right)>0\;\mathrm{and}\;x^{2}-y^{2}>6\\
\Sigma_{+-}^{\left[1\right]} & 1-6r^{2}+3r^{4}+2\cos\left(2\theta\right)>0\;\mathrm{and}\;x^{2}-y^{2}<6
\end{cases}\vspace{2mm}\\
\Sigma_{e}^{IV}(z)&=&\Sigma_{--}^{\left[0\right]}\vspace{2mm}\\
\Sigma_{e}^{V}(z)&=&
\begin{cases}
\Sigma_{+-}^{\left[0\right]} & 1-6r^{2}+3r^{4}+2\cos\left(2\theta\right)<0\\
\Sigma_{+-}^{\left[1\right]} & 1-6r^{2}+3r^{4}+2\cos\left(2\theta\right)>0\;\mathrm{and}\;x^{2}-y^{2}>6\\
\Sigma_{+-}^{\left[-1\right]} & 1-6r^{2}+3r^{4}+2\cos\left(2\theta\right)>0\;\mathrm{and}\;x^{2}-y^{2}<6
\end{cases}\vspace{2mm}\\
\Sigma_{e}^{VI}(z)&=&
\begin{cases}
\Sigma_{--}^{\left[0\right]} & x^{2}-y^{2}>6\\
\Sigma_{--}^{\left[-1\right]} & x^{2}-y^{2}<6
\end{cases}
\end{array}
\vspace{1mm}
\end{equation}
where $x\equiv \mathrm{Re}\,z , y\equiv \mathrm{Im}\,z , r\equiv\left|\xi_{+-}\right|$ and $\theta\equiv \mathrm{Arg}\left(\xi_{+-}\right)\in\left(-\pi,\pi\right]$. 
Finally, we define our “physical” self-energy, $\Sigma_{e}^{phy}\left(z\right)$, as a piecewise single valued function whose domain covers the whole complex plane except for some possible complex lines where discontinuities in the real and imaginary parts of $\Sigma_{e}^{phy}\left(z\right)$ are found. These special regions must be taken into account when performing the detour integration given by Eq.~\eqref{eq:Propagator inversion}.
\begin{equation}
\Sigma_{e}^{phy}\left(z\right)=
\begin{cases}
\begin{array}{llcl}
\Sigma_{e}^{I}(z) & \left|x\right|>2\sqrt{3} & or & y\geq0\\
\Sigma_{e}^{II}(z) & x\in\left(-2\sqrt{3},-2\sqrt{2}\right) & and & y<0\\
\Sigma_{e}^{III}(z) & x\in\left(-2\sqrt{2},-2\right) & and & y<0\\
\Sigma_{e}^{IV}(z) & x\in\left(-2,2\right) & and & y<0\\
\Sigma_{e}^{V}(z) & x\in\left(2,2\sqrt{2}\right) & and & y<0\\
\Sigma_{e}^{VI}(z) & x\in\left(2\sqrt{2},2\sqrt{3}\right) & and & y<0
\end{array}
\end{cases}
\vspace{1mm}
\end{equation}
Consequently, the corresponding “physical” propagator reads:
\begin{equation}
G_{e}^{phy}(z)=\frac{1}{z-\Delta-\Sigma_{e}^{phy}(z)}
\end{equation}

\end{appendices}

\bibliography{articles,articles2,books}

\end{document}